\documentclass[aps,pra,reprint,superscriptaddress,longbibliography]{revtex4-1}
\usepackage[utf8]{inputenc}
\usepackage[T1]{fontenc}

\usepackage{amsmath,amssymb}
\usepackage{graphicx}
\usepackage{grffile}
\usepackage{microtype}
\usepackage[usenames,dvipsnames]{xcolor}
\usepackage[colorlinks,linkcolor=blue,citecolor=blue,urlcolor=blue]{hyperref}

\def\equationautorefname~#1\null{Eq.~(#1)\null}

\newcommand{\appref}[1]{\href{#1}{\appendixautorefname~\ref*{#1}}}

\DeclareMathOperator{\Tr}{Tr}

\begin{document}

\title{Long-distance heat transfer between molecular systems through a hybrid plasmonic-photonic nanoresonator}

\author{S. Mahmoud Ashrafi}
\affiliation{Department of Physics, Tarbiat Modares University, Tehran, Iran}
\author{R. Malekfar}
\affiliation{Department of Physics, Tarbiat Modares University, Tehran, Iran}
\author{A. R. Bahrampour}
\affiliation{Department of Physics, Sharif University of Technology, Tehran, Iran}
\author{Johannes Feist}
\email{johannes.feist@uam.es}
\affiliation{Departamento de Física Teórica de la Materia Condensada and
Condensed Matter Physics Center (IFIMAC), Universidad Autónoma de Madrid,
E-28049 Madrid, Spain}

\begin{abstract}
We introduce a hybrid plasmonic-photonic cavity setup that can be used to induce
and control long-distance heat transfer between molecular systems through
optomechanical interactions. The structure consists of two separated plasmonic
nanoantennas coupled to a dielectric cavity. The hybrid modes of this resonator
can combine the large optomechanical coupling of the sub-wavelength plasmonic
modes with the large quality factor and delocalized character of the cavity mode
that extends over a large distance ($\sim\mu$m). We show that this can lead to
effective long-range heat transport between molecular vibrations that can be
actively controlled through an external driving laser.
\end{abstract}

\maketitle

\section{Introduction}
Energy transfer between quantum emitters (quantum dots, molecules, atoms,
\ldots) is a process of fundamental importance for a large range of phenomena in
quantum information, quantum thermodynamics, quantum biology, photosynthesis,
solar cells, etc.~\cite{Nagali2009, Northup2014, Nalbach2010, Dubi2011,
Katz2016, Lee2007, Scholes2011, High2008, Menke2013}. One powerful strategy to
modify these processes is by coupling the emitters with an electromagnetic mode
and mediating transport through photon absorption and
emission~\cite{Gerry2004,Messina2012,Feist2015,Schachenmayer2015}. Since the
light-matter coupling strength is inversely proportional to the effective mode
volume, the largest single-emitter coupling strengths up to now have been
reached with nanoplasmonic resonators due to their ability to confine light in
ultra-small sub-wavelength volumes down to $V_{p} \approx 10^{-8}
\lambda^{3}$~\cite{Kim2015,Chikkaraddy2016,Ojambati2019}, where $\lambda$ is the
free-space wavelength. These capabilities have led to a large range of
applications of nanoplasmonic systems~\cite{Fernandez-Dominguez2017}, including
in the context of quantum electrodynamics (QED) and quantum optics, such as for
strong coupling with a single molecule at room
temperature~\cite{Chikkaraddy2016,Ojambati2019} or single-photon
sources~\cite{Hoang2016, Straubel2017}.

Over the last few years, it has also been realized that the process of
surface-enhanced Raman scattering, which is a well-known strategy for enhancing
the Raman scattering signal of molecules by many orders of
magnitude~\cite{Kneipp2006,Schlucker2014}, can alternatively be interpreted
through the framework of quantum optomechanics~\cite{Roelli2016,
Schmidt2016Quantum}. In this context, we have recently shown that a localized
surface plasmon resonance (LSPR) modes can mediate heat transfer between two
molecules through its optomechanical coupling to the vibrations in each
molecule~\cite{Ashrafi2019}. However, this approach is limited in practice due
to the required small (nm-scale) distance between molecules coupled to the same
sub-wavelength LSPR mode, and additionally suffers from the low quality factors
($Q \sim 10$) of plasmonic resonators that are unavoidable due to the large
intrinsic losses of the metals providing the sub-wavelength confinement. Hybrid
plasmonic-photonic (i.e., metallo-dielectric) cavities present an intriguing
possibility to circumvent these limitations by combining the strong light-matter
interactions of plasmonic systems with the possible large lifetimes (high
quality factors) of dielectric structures. Such approaches have been shown to
improve the performance of existing applications and allow novel applications
for a broad range of examples such as strong light-matter
coupling~\cite{Xiao2012,Conteduca2017,Gurlek2018}, optical
trapping~\cite{Mossayebi2016}, surface-enhanced Raman
scattering~\cite{Peyskens2016}, label-free detection of
molecules~\cite{DeAngelis2008, Dantham2013}, biosensing~\cite{Conteduca2016},
optoplasmonic sensors~\cite{Xavier2018}, or refractometers and nanoparticle
trapping~\cite{Hu2013Hybrid}.

\begin{figure}
  \includegraphics[width=\linewidth]{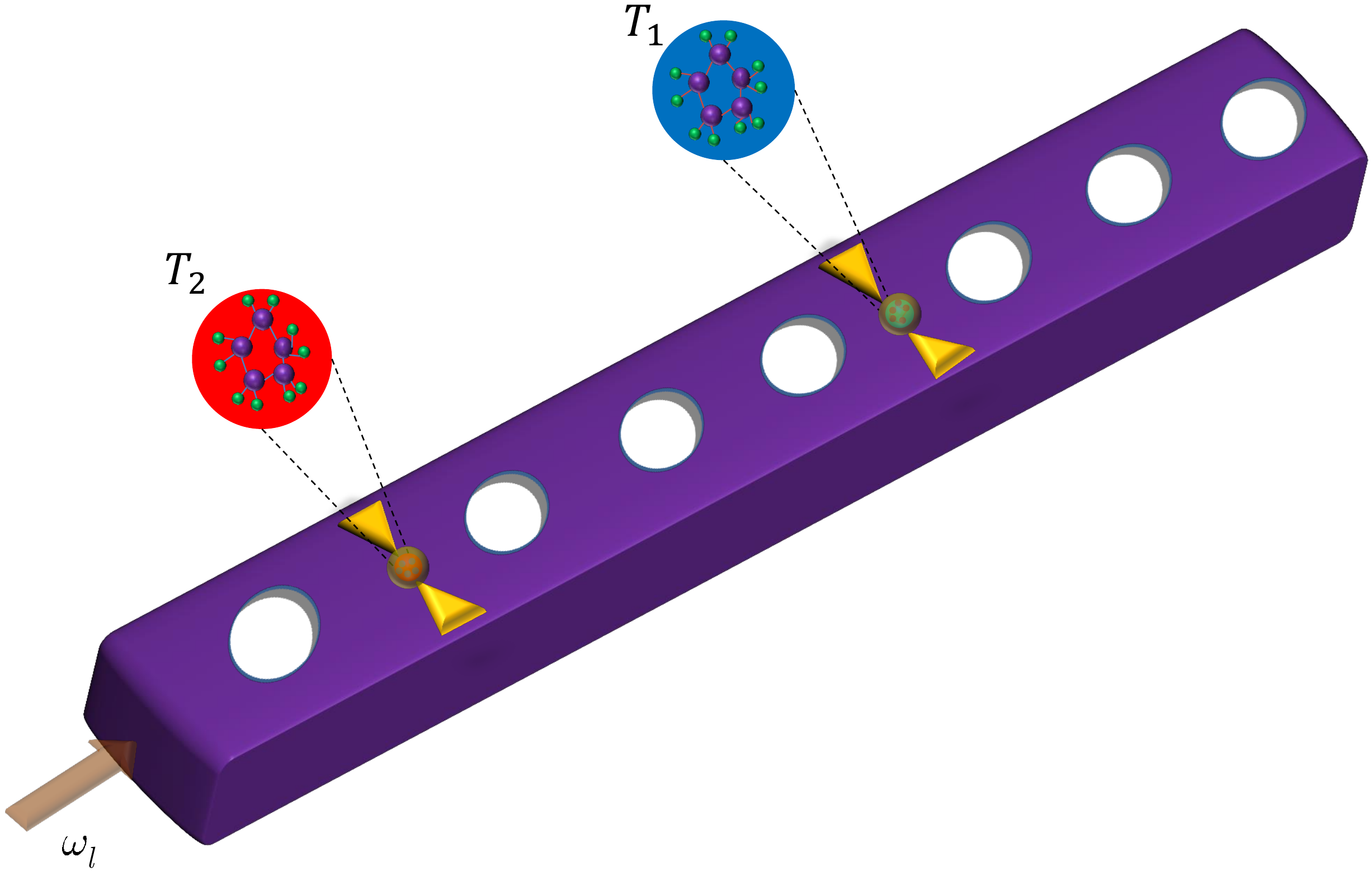}
  \caption{Sketch of the hybrid plasmonic-photonic nanoresonator containing
  two separated molecules placed in the hot spots of two bowtie nanoantennas
  with different temperatures (the red circle displays a ``hot'' molecule and
  the blue circle displays a ``cold'' molecule).}
  \label{fig:1}
\end{figure}

In this article, we show that a hybrid photonic-plasmonic cavity can enable
efficient long-range heat transfer between spatially separated molecules through
optomechanical interactions. The setup we propose, sketched in \autoref{fig:1},
consists of two metallic nanoantennas coupled to a photonic cavity such as a
photonic crystal (PC) beam or a dielectric mirror cavity. Such a system supports
two different kinds of electromagnetic modes, strongly localized LSPR modes on
each nanoantenna, and an extended cavity mode, which we assume to be driven by
an external laser. The maximum distance between the nanoantennas is determined
by the extension of the cavity mode and can be on the order of (several times)
the laser wavelength. A molecule is placed in the hot spot of each nanoantenna,
with each molecule coupled to a local heat bath at temperature $T_1$ (cold) and
$T_2$ (hot), respectively. As we will show below, such a system can be used to
efficiently transfer heat between the molecular vibrational modes, with the
driving laser providing active control over the heat transfer process.

The paper is organized as follows: in section II, we introduce the quantum
optomechanical model we use, both in terms of the original uncoupled (LSPR and
PC) modes as well as in terms of hybrid modes obtained from their coupling. In
section III, we present the main results, including a discussion of the
influence of the main parameters of the system, and in particular show the
consequence of coupling between vibrational modes of different frequencies. We
conclude with a section summarizing and discussing the results.

\section{Theoretical model and framework}

\begin{figure}
  \includegraphics[width=0.8\linewidth]{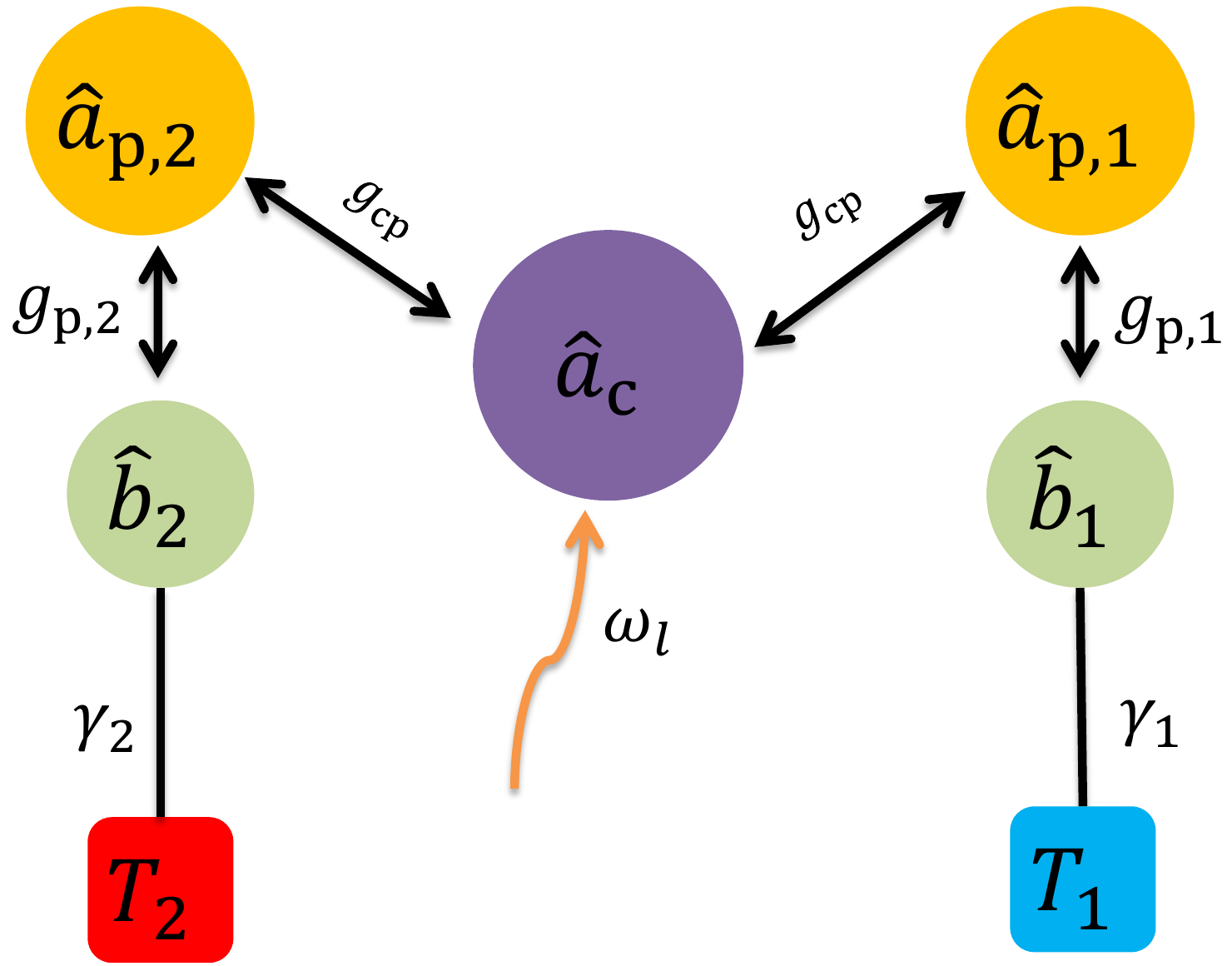}
  \caption{Scheme of the model with all relevant parameters.}
  \label{fig:2}
\end{figure}

Our model, schematically depicted in \autoref{fig:2}, treats three photonic
modes and two molecules. We note that while the direct quantization of hybrid
cavity modes is far from trivial~\cite{Hughes2018,Franke2019}, we have recently
shown that they can be represented accurately within a description containing a
few interacting modes (without counterrotating terms) that each decay
independently~\cite{Medina2020}. The model we use in the following is based on
such a description. The three photonic modes are given by one cavity mode, with
frequency $\omega_c$ and bosonic annihilation operator $a_c$, and two LSPRs,
with frequencies $\omega_{p,i}$ and bosonic annihilation operators $a_{p,i}$ ($i
= 1,2$). One molecule is assumed to be placed in the hot spot of each LSPR, with
each molecule represented by a single vibrational mode. These modes are
approximated as harmonic oscillators, with frequencies $\nu_{j}$ and
annihilation operators $b_j$ ($j = 1,2$). We assume that an external laser with
frequency $\omega_l$ and amplitude $\Omega$ drives the cavity mode. The total
Hamiltonian of the system in the rotating frame of the laser and within the
rotating wave approximation can then be written as (here and in the following,
we set $\hbar=1$)
\begin{subequations}\label{eq:Hamiltonian}
\begin{align}
  H &= H_\mathrm{ph} + H_\mathrm{m} + H_\mathrm{I} + H_\mathrm{d},\\
  H_\mathrm{ph} &= \sum_\alpha \delta_\alpha a_\alpha^\dagger a_\alpha + \sum_{i=1}^2 g_\mathrm{cp} (a_c^\dagger a_{p,i} + a_c a_{p,i}^\dagger),\\
  H_\mathrm{m} &= \sum_{j=1}^2 \nu_j b_j^\dagger b_j,\\
  H_\mathrm{I} &= -\sum_{\alpha,j} g_{\alpha j} a_{\alpha}^{\dagger} a_{\alpha} (b_{j}^{\dagger} + b_{j})\\
  H_\mathrm{d} &= -i\Omega (a_c^\dagger - a_c)
\end{align}
\end{subequations}
where $\delta_\alpha = \omega_\alpha - \omega_l$, the sum over $\alpha$ includes
the three photonic modes and $g_{cp}$ accounts for the interaction between the
cavity mode and the LSPRs, which depends on the placement of the nanoantennas
relative to the optical cavity~\cite{KamandarDezfouli2017}. Here and in the
following, we consider two identical plasmonic cavities placed at equivalent
positions sufficiently separated from each other that direct plasmon-plasmon
interactions are negligible. The interaction between photonic modes and
vibrations, $H_\mathrm{I}$ is approximated through their optomechanical
coupling, which is justified when $\omega_{\alpha} \gg
\nu_{j}$~\cite{Roelli2016,Schmidt2016Quantum,Schmidt2017}. The optomechanical
coupling constants are given by $g_{\alpha j} = \frac{\omega_{\alpha} Q_{j}^{0}
R_{j}}{2\varepsilon\varepsilon_{0}V_{\alpha}}$, where $Q_{j}^{0}$ is the zero
point amplitude of vibration $j$, $R_{j}$ is its Raman activity,
$\varepsilon$ is the relative permittivity of the surrounding medium and
$V_{\alpha}$ is the effective mode volume of the photonic mode. Note that the
``bare'' photonic mode volume is large ($V \sim \lambda^3 \sim 10^{9}~$nm$^3$), so that its
direct optomechanical coupling to the molecules is negligible, and that each
plasmonic antenna only interacts with its ``own'' molecule, $g_{p_1 2} =
g_{p_2 1} = 0$.

In addition to the coherent dynamics described by the Hamiltonian, we also model
the incoherent dynamics due to the interaction between the different components
of the system and their environment. The dynamics of the system density matrix
is described within the Lindblad master equation formalism~\cite{Breuer2007}:
\begin{equation}
  \frac{\mathrm{d}\rho}{\mathrm{d}t} = -i [H,\rho] + \sum_\alpha L_{a_\alpha}[\rho] + \sum_j L_{b_{j}}[\rho],
\end{equation}
with 
\begin{align}
  L_{a_{\alpha}}[\rho] &= \kappa_\alpha D_{a_{\alpha}}[\rho],\\
  L_{b_{j}}[\rho] &= \gamma_{j} (\bar{n}_{j} + 1) D_{b_j}[\rho] + \gamma_{j} \bar{n}_{j} D_{b_{j}^{\dagger}}[\rho].
\end{align}
Here, $\kappa_\alpha$ is the decay rate of the photonic mode $\alpha$,
$\gamma_j$ is the molecular damping rate of molecule $j$, and $\bar{n}_{j} =
1/\left[\exp\left(\frac{\nu_j}{k_B T_j}\right) - 1\right]$ is the mean phonon
number of molecule $j$ when it is in thermal equilibrium with its associated
bath at temperature $T_j$. Finally, $D_{A}[\rho]$ is a Lindblad dissipator,
\begin{equation}
  D_{A}[\rho] = A \rho A^{\dagger} - \frac{1}{2} \{ A^{\dagger}A, \rho \}.
\end{equation} 

In order to quantify the heat transfer between the molecules, we follow the same
approach as in our previous paper~\cite{Ashrafi2019}, which we summarize in the
following: We first solve for the steady state of the system,
$\frac{\mathrm{d}\rho_\mathrm{ss}}{\mathrm{d}t} = 0$, and then extract an
effective temperature for the molecular vibrations,
\begin{equation}
  T_{j}^{\text{eff}} = \frac{\nu_{j}}{k_{B}\log(1 + 1/n_{j})},
\end{equation}
based on the average phonon number $n_{j} = \langle b_{j}^{\dagger} b_{j}\rangle
= \Tr(b_{j}^{\dagger} b_{j} \rho_\mathrm{ss})$. For this effective temperature
to correspond to a physical temperature, the population of the energy levels
should again follow a thermal distribution. We have checked for all the results
presented below that this is indeed the case, i.e., that the steady-state
distributions of the phonon populations are well approximated by thermal
Boltzmann distributions, and the effective temperatures obtained can thus indeed
be interpreted as the steady-state physical temperatures of the respective
vibrational modes.

All the numerical results have been obtained using the QuTiP
package~\cite{Johansson2012,Johansson2013}, and plots have been prepared using
matplotlib~\cite{Hunter2007,Caswell2020}. We note that in contrast to our
previous work~\cite{Ashrafi2019}, which used only a single photonic mode, we
here treat a relatively big quantum system consisting of five bosonic modes
(three photonic and two vibrational) within a density matrix description, such
that the numerical size of the density matrix is $5^N \times 5^N$ if $N$ states
are used for each mode. The size of the (numerically sparse) Liouvillian
superoperator is the square of that number. In order to obtain numerical
convergence while keeping the size of the calculations manageable, we use a
basis in which the total number of excitations (photons + phonons) is
restricted, instead of using a cutoff for each mode separately. The results
presented below use a maximum number of five excitations, which we have checked
to give converged results for the parameters used.

\section{Results and Discussion}

\begin{figure}
  \includegraphics[width=\linewidth]{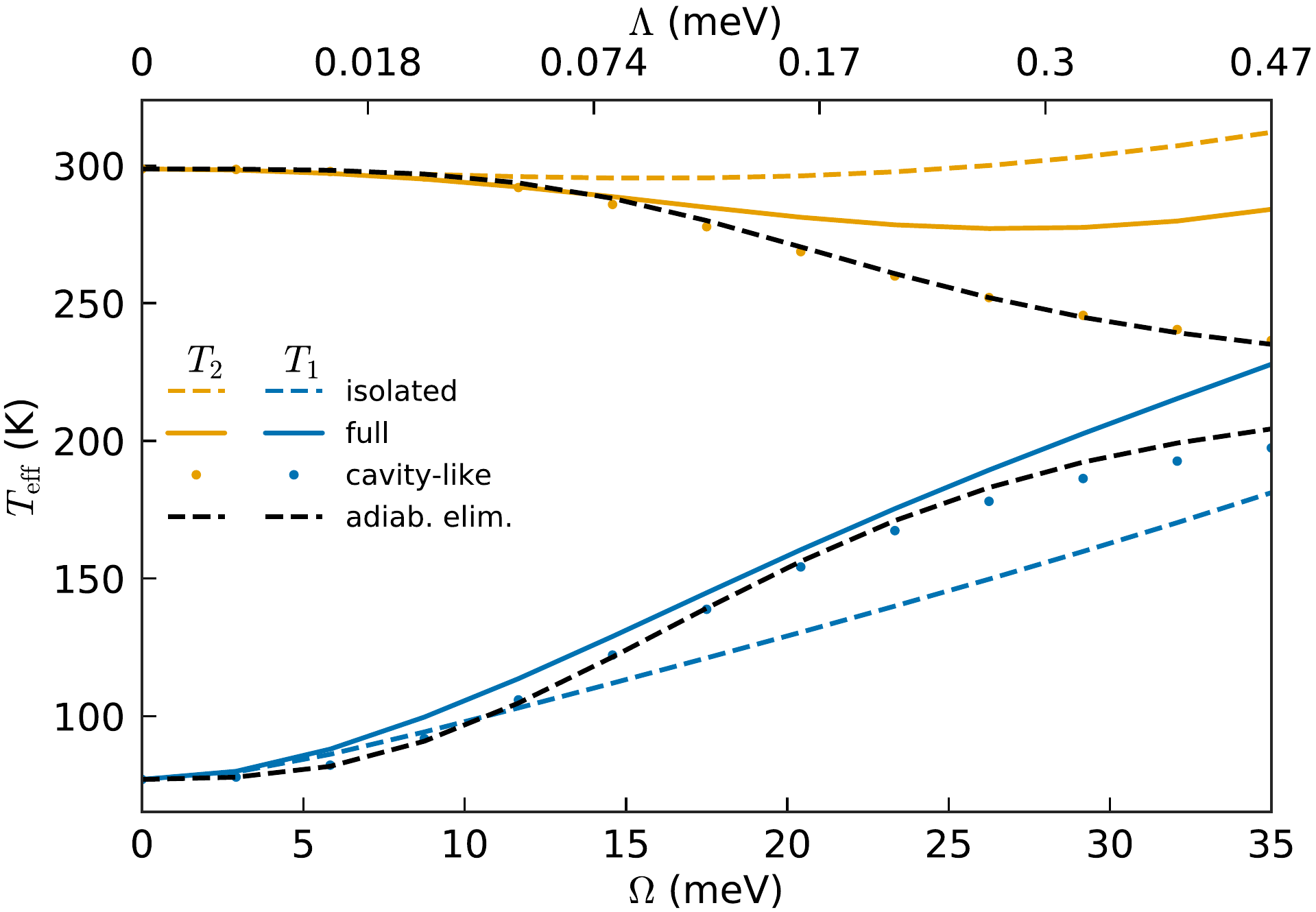}
  \caption{Effective temperature of two identical molecules coupled to heat
  baths at different temperatures, and indirectly coupled to each other through
  optomechanical interactions with hybrid cavity modes. The full results (solid
  lines) are compared to the case of each molecule in isolation (dashed lines),
  and to the result when only the cavity-like hybrid mode is taken into account
  (dots) and when that mode is adiabatically eliminated (dashed black lines).
  Parameters are given in the main text.}
  \label{fig:3}
\end{figure}

We now investigate long-range energy transfer between two molecules, each
coupled to a plasmonic nanoantenna with a low-Q resonance, with the two antennas
coupled to the same high-Q cavity mode. The model we use is similar to the one
proposed by Kamandar Dezfouli and Hughes~\cite{KamandarDezfouli2017}, with the
addition of a second plasmonic antenna and molecule. It consists of a photonic
crystal supporting a mode with frequency $\omega_{c} = 1.61$~eV and a high
quality factor, $Q_{c} = 3 \times 10^{5}$ (i.e. $\kappa_{c} = 5.4 \times
10^{-2}$~meV), and two identical bowtie nanoantennas separated by a relatively
large distance ($\sim\mu$m), with a dipolar LSPR at frequency $\omega_{p,1} =
\omega_{p,2} = 2.2$~eV, and relatively low quality factor of $Q_{p,1} = Q_{p,2}
= 40$ (i.e., $\kappa_{p,1} = \kappa_{p,2} = 55$~meV). One molecule is placed in
the hot spot (i.e., the center) of each nanoantenna. We first treat the case of
two identical molecules, with a vibrational mode with angular frequency $\nu_{1}
= \nu_{2} = 40$~meV, and vibrational damping rate of $\gamma_{1} = \gamma_{2} =
0.1$~meV. Each molecule is coupled to a local heat bath at a different
temperature, where here and in the following we set $T_{1} = 77$~K, $T_{2} =
300$~K. The optomechanical coupling rate for the LSPR modes is taken as $g_{p} =
75$~meV, corresponding to close-to-resonant Raman
transitions~\cite{KamandarDezfouli2017}, while the direct optomechanical
coupling to the cavity mode is so small as to have negligible effect (due to the
large mode volume). However, the coupling between the cavity mode and each LSPR,
with coupling constant $g_{cp} = 200$~meV, mediates an indirect coupling between
the LSPR modes, and thus between the molecules.

\begin{figure*}
  \includegraphics[width=\linewidth]{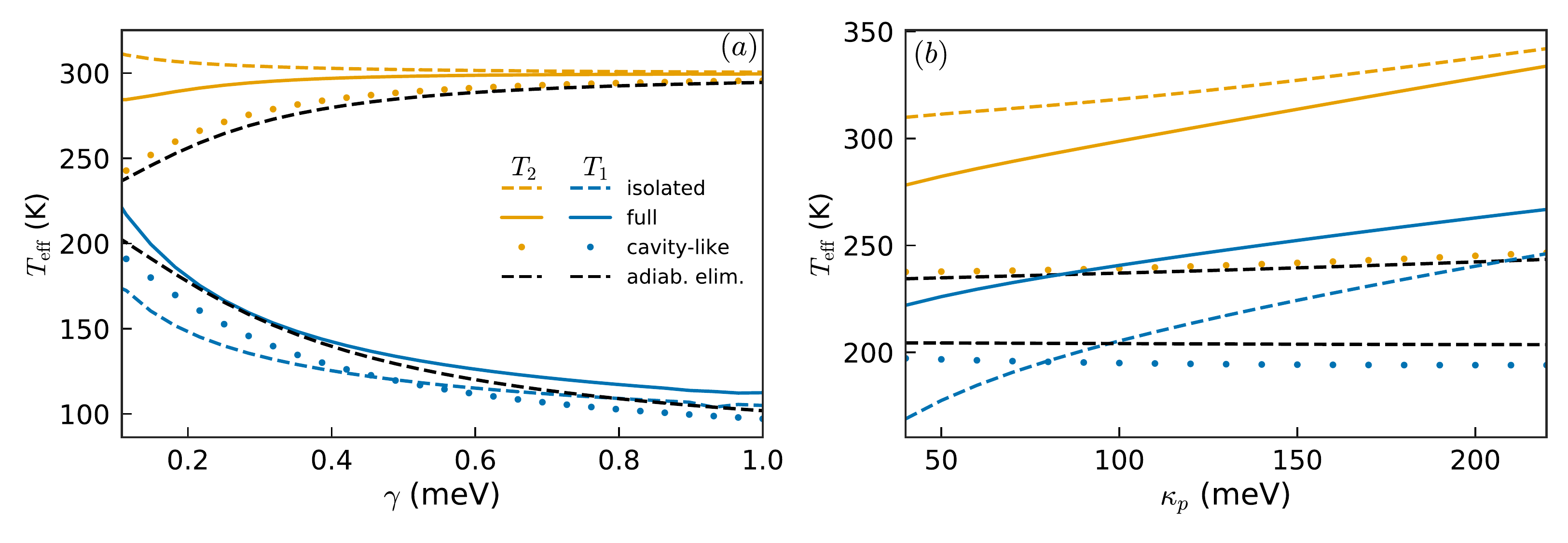}
  \caption{Effective temperature of molecules as a function of (a) molecular
  damping rate and (b) plasmonic damping rate. The external driving rate is
  $\Omega=35$~meV, with all other parameters as in \autoref{fig:3}.}
  \label{fig:4}
\end{figure*}

We focus on the case where the external laser drives the cavity mode with red
detuning, $\omega_l = 1.41$~eV. \autoref{fig:3} shows the effective
temperature of each molecule as a function of the external laser driving
strength $\Omega$. As expected, for $\Omega=0$, the photonic modes have no
influence on the molecules, and each molecular vibration is in equilibrium with
its local heat bath. When the laser is turned on, there are two effects on the
molecular system. First, the optomechanical coupling for each molecule in
isolation leads to heating/cooling of the vibrations through Stokes/anti-Stokes
transitions, such that they are driven out of equilibrium with their respective
heat baths. This well-known single-molecule effect~\cite{Schmidt2017} is shown
in dashed lines in \autoref{fig:3}. For the parameters considered here, this
results in heating for both molecules, significant for the colder molecule
($T_{1}^{\text{eff}}$ increases from $77$~K to $\approx\!170$~K), and less
pronounced for the hotter molecule ($T_{2}^{\text{eff}}$ increases from $300$~K
to $\approx\!315$~K). Second, there is an effective molecule-molecule coupling
mediated by the photonic modes (through the successive couplings
molecule-LSPR-cavity-LSPR-molecule), which enables energy transfer between the
molecules, the efficiency of which can be estimated from the effective molecular
temperatures (solid lines in \autoref{fig:3}) compared to the single-molecule
cases. As seen in \autoref{fig:3}, this can be a significant effect, enabling
efficient transfer of energy between the molecules under external pumping of the
high-Q cavity mode. For example, for $\Omega = 35~\text{meV}$, we obtain
$T_{1}^{\text{eff}} \approx 220$~K, $T_{2}^{\text{eff}} \approx 280$~K, showing
that the hot molecule is efficiently cooled through long-distance heat transfer
to the cooler molecule, which is significantly heated in return. The proposed
setup could thus indeed be used for externally controlling the heat or energy
transfer between molecules that are spatially separated by significant
distances. The pumping rate is the main parameter controlling this effect, and
increasing it improves the energy exchange, causing the effective temperatures
to approach each other.

Additional insight into the long-range heat transfer can be obtained by
considering a simplified model where the three photonic modes are diagonalized
to obtain hybrid modes (see \appref{sec:hybrid_framework} for details). In this
case, one obtains one high-Q and two lower-Q hybrid modes (one of which is a
``dark'' plasmonic mode consisting of the antisymmetric combination of plasmonic
antenna modes and does not hybridize with the cavity-like mode). The high-Q mode
is dominated by the cavity mode contribution. We thus label it as the
``cavity-like'' mode in the following. Considering only the optomechanical
interaction with the cavity-like mode (dots in \autoref{fig:3}) reproduces the
main effect of optomechanically induced heat transfer, although it
underestimates the laser-induced heating, in particular of the hotter molecule.
Adiabatically eliminating the cavity-like mode within this single-mode
approximation then gives an effective molecule-molecule coupling term $\Lambda
(b_1^\dagger b_2 + b_1 b_2^\dagger)$~\cite{Ashrafi2019}. The effective
temperatures obtained within this approximation (dashed black lines in
\autoref{fig:3}, with the corresponding value of $\Lambda$ shown in the upper
axis) reproduce the single-mode results almost perfectly.

We next treat the influence of other parameters on long-distance heat flow
between the molecules. We show the effective temperature of the molecules as a
function of the coupling rate $\gamma = \gamma_{1} = \gamma_{2}$ between each
vibration and its local heat bath in \autoref{fig:4}(a). This rate determines
the rate of heat exchange with the local bath, and the temperatures of the
vibrational modes thus approach the bath temperatures as $\gamma$ is increased.
For the ``hot'' molecule 2, this is enough to make the temperature of the
molecule approach its undriven value ($T_2 = 300$~K). While the effect is also
very noticeable for the ``cold'' molecule, with a decrease in temperature from
$\approx\!220$~K to $\approx\!110$~K as $\gamma$ is increased from $0.1$~meV to
$1$~meV, it is not enough to completely counteract the externally induced heat
transport. Again, the single-mode approximation only taking into account the
cavity-like hybrid mode reproduces the observed trends quite well while
underestimating the laser-induced heating of the hotter molecule.

We additionally investigate the effect of the photonic system parameters on the
heat transfer effect. In particular, as shown in \autoref{fig:4}(b), an increase
in the plasmonic cavity loss $\kappa_{p} = \kappa_{p,1} = \kappa_{p,2}$ leads to
a reduction of heat transfer, while simultaneously inducing more heating on the
single-molecule level. As the single-mode approximation that only takes into
account the high-Q cavity-like mode underestimates the plasmon-induced heating,
it is not capable of reproducing this trend accurately, and the effective
temperatures within this approximation are approximately constant as a function
of $\kappa_p$, see the dots in \autoref{fig:4}(b). The effect seen here shows
that it is desireable to use plasmonic nanoantennas of the highest quality
possible to minimize direct optomechanical heating while simultaneously reaching
large effective optomechanical coupling strengths. Similarly, we find that
improving the quality factor of the delocalized cavity mode (i.e., decreasing
its loss rate) also leads to improved heat transfer (not shown).

\begin{figure}[t]
  \includegraphics[width=\linewidth]{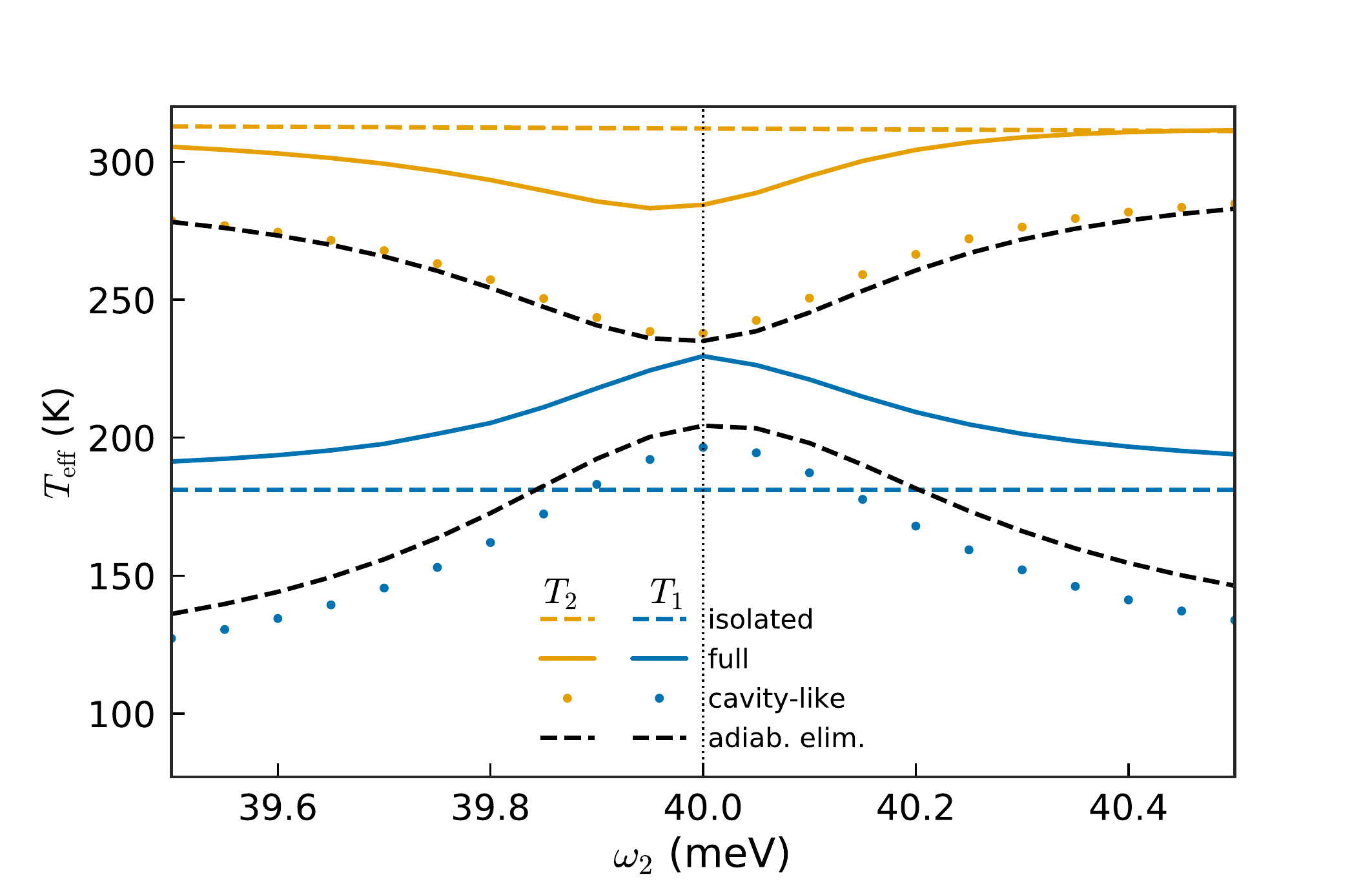}
  \caption{Heat transfer process in the non-symmetric case as a function of the
  vibrational frequency of the ``hot'' molecule $2$.}
  \label{fig:5}
\end{figure}

\begin{figure*}[tb]
  \includegraphics[width=\linewidth]{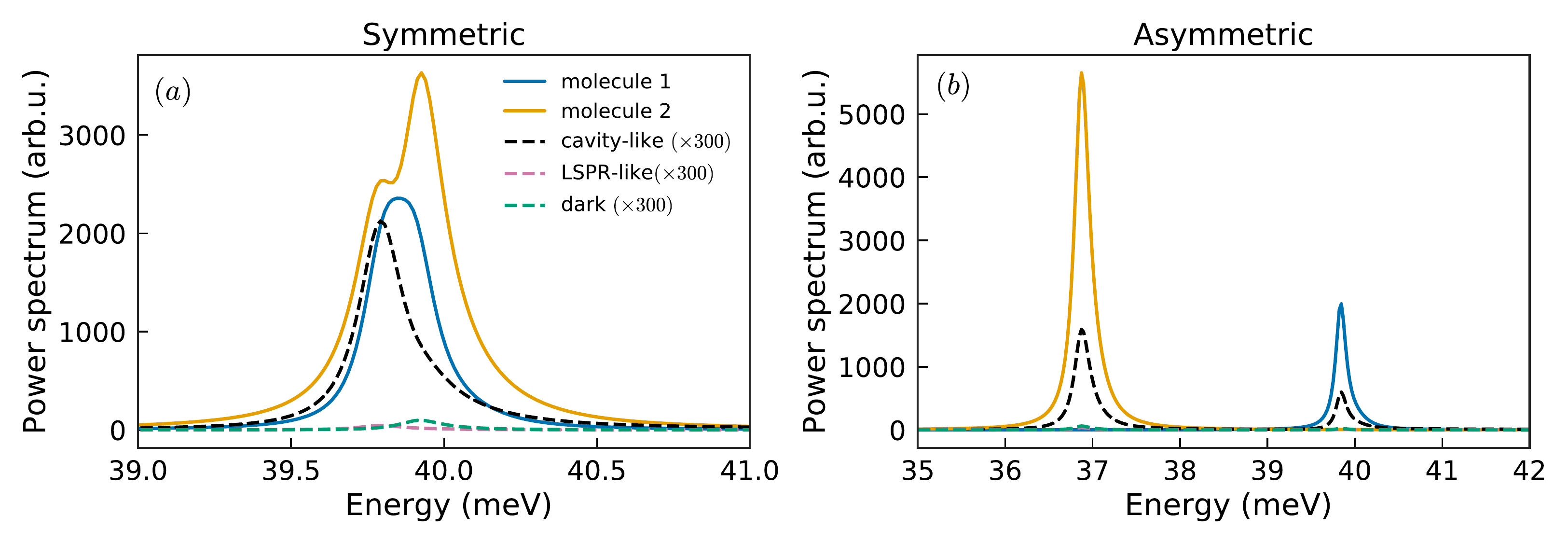}
  \caption{(a) Power spectral density in symmetric system for hot molecule (red
  solid line), cold molecule (blue solid line) and the PC mode (orange dashed).
  (b) The PSD for non-symmetric system when $\omega_{1} = 50~$meV, $\omega_{2} =
  47~$meV.}
  \label{fig:6}
\end{figure*}

We now treat the non-symmetric case where the vibrational frequencies of the two
molecules are not identical (while keeping all other parameters fixed for
simplicity), i.e., we explore whether it is possible to enable heat transfer
between molecules even if their vibrational frequencies are not equal. In
\autoref{fig:5}, we plot the effective vibrational temperatures of the two
molecules when the vibrational frequency $\nu_2$ of the ``hot'' molecule is
changed, with $\nu_1$ fixed at $40$~meV and all other parameters as in
\autoref{fig:3}. We find relatively efficient heat transfer only when the two
vibrations are close to resonance ($\nu_1\approx\nu_2$). These results are
consistent with the case where two molecules are coupled to the \emph{same}
plasmonic nanocavity mode~\cite{Ashrafi2019}, and imply that a
single-vibrational-mode approximation is justified.

To gain a better understanding of this effect, we next analyze the power
spectral density (PSD) of the involved modes, which measures their oscillation
spectrum as a function of frequency~\cite{Gardiner2004}. The PSD for a mode with
annihilation operator $A$ is given by the Fourier transform of the correlation
spectrum of the operator in the steady state, 
\begin{equation}
  S_{A}(\omega) = \int e^{- i\omega t} \left\langle A^{\dagger}(t)A(0) \right\rangle_{\text{ss}}\mathrm{d}\omega.
\end{equation}
The PSD for the symmetric case of identical molecular vibrations is shown in
\autoref{fig:6}(a). For the photonic mode basis, we here use the hybrid photonic
modes obtained after diagonalizing the coupled cavity and LSPR modes. This
figure shows that the cavity-like hybrid mode (black dashed line) is the
principal photonic mode involved in the dynamics, with a modulation imprinted on
it at frequencies close to the vibrational ones. As discussed previously, this
mode mediates the effective long-range molecule--molecule coupling. In the
current regime, the effective linewidth of the PSD of the cavity-like mode is
\emph{not} determined by its loss rate ($\kappa_- \approx 8$~meV), but much
closer to the molecular vibrational linewidth. When the two molecular vibrations
are detuned from each other, shown in \autoref{fig:6}(b) for vibrational
frequencies $\nu_{1} = 40$~meV and $\nu_{2} = 37$~meV, each molecule only
oscillates at its native frequency, and the cavity-like mode has two essentially
independent peaks at $\nu_1$ and $\nu_2$. The off-resonant fluctuations induced
on the cavity-like mode by the two vibrational modes then prevent effective
long-range energy transfer between them.

\section{Summary \& Conclusions:}

To summarize, we have investigated the optomechanical heat transfer mechanism
between two spatially separated molecules at different local temperatures placed
in a hybrid plasmonic-photonic nanoresonator. The hybrid cavity considered
consists of two plasmonic nanoantennas supporting LSPR modes that are both
coupled to the same high-quality cavity mode (such as supported by a photonic
crystal cavity or a Fabry-Perot cavity with highly reflective mirrors). The
cavity mode is driven by an external red-detuned laser. We have shown that in
such a hybrid setup, the cavity-like mode, which itself has negligible
optomechanical coupling to the molecular vibrations, can behave like a mediator
that provides an effective molecule--molecule coupling over large distances of
the order of several free-space wavelengths, while the strongly sub-wavelength
plasmonic modes provide the required large optomechanical coupling strengths.
The hybrid setup considered can thus induce long-range heat transfer between
molecules. Additionally, the heat transfer is fully controlled by an external
driving laser, and can thus be dynamically turned off or on. Furthermore, we
have found that this mechanism only occurs for close-to-resonant vibrational
modes, which can be understood by considering the molecule-induced oscillation
of the hybrid cavity modes through their power spectral density.

\begin{acknowledgements}
This work has been funded by the European Research Council through grant
ERC-2016-StG-714870 and by the Spanish Ministry for Science, Innovation, and
Universities -- Agencia Estatal de Investigación through grants
RTI2018-099737-B-I00, PCI2018-093145 (through the QuantERA program of the
European Commission), and MDM-2014-0377 (through the María de Maeztu program for
Units of Excellence in R\&D), as well as through a Ramón y Cajal grant (JF) and
support from the Iranian Ministry of Science, Research and Technology (SMA).
\end{acknowledgements}

\appendix

\section{Hybrid modes}\label{sec:hybrid_framework}

We here discuss the properties of the hybrid modes obtained by diagonalizing the
``photonic'' subsystem consisting of the cavity mode and two localized surface
plasmon resonances. The photonic Hamiltonian can be written as $H_{\mathrm{ph}}
= \vec{A}^\dagger \boldsymbol{\mathcal{H}} \vec{A}$, where $\vec{A} =
(a_c,a_{p,1},a_{p,2})^T$ collects the photonic annihilation operators and the
matrix $\mathcal{H}$ is
\begin{equation}
  \boldsymbol{\mathcal{H}} = \begin{pmatrix}
      \delta_{p} & 0 & g_{cp} \\
      0 & \delta_{p} & g_{cp} \\
      g_{cp} & g_{cp} & \delta_{c}
    \end{pmatrix}.
\end{equation}
Diagonalization of $\boldsymbol{\mathcal{H}}$ gives the hybrid mode energies
\begin{align}
  \delta_{\pm} &= \frac{{\delta_{p} + \delta}_{c}}{2} \pm \frac12 \sqrt{(\delta_{p} - \delta_{c})^{2} + 8g_{cp}^{2}},\\
  \delta_{D} &= \delta_{p},
\end{align}
and annihilation operators
\begin{align}
  a_{-} &=  \cos\theta a_{c} + \frac{\sin\theta}{\sqrt{2}} (a_{p,1} + a_{p,2}),\\
  a_{+} &= -\sin\theta a_{c} + \frac{\cos\theta}{\sqrt{2}} (a_{p,1} + a_{p,2}),\\
  a_{D} &= \frac{1}{\sqrt{2}} (a_{p,1} - a_{p,2}),
\end{align}
where $\tan2\theta = \sqrt{8} g_{cp} / (\delta_{p} - \delta_{c})$. If the
coupling $g_{cp}$ is not too large compared to the detuning $\delta_{p} -
\delta_{c}$, the hybrid modes are a `small' rotation of the original basis and
can be identified as a cavity-like hybrid mode ($a_-$), an LSPR-like hybrid mode
($a_+$), and an LSPR dark mode $a_D$ that has no cavity contribution. The
original operators can be represented in the new basis as
\begin{align}
  a_{c} &= \cos\theta a_{-} - \sin\theta a_{+},\\
  a_{p,1} &= \frac{1}{\sqrt{2}}(\sin\theta a_{-} + \cos\theta a_{+} + a_{D}),\\
  a_{p,2} &= \frac{1}{\sqrt{2}}(\sin\theta a_{-} + \cos\theta a_{+} - a_{D}).
\end{align}
The changed parts of the Hamiltonian in the new hybrid basis are given by
\begin{subequations}\label{eq:Hamiltonian_hybrid}
\begin{align}
    H_\mathrm{ph} &= \sum_\beta \delta_\beta a_\beta^\dagger a_\beta,\\
    H_\mathrm{I} &= -\sum_{\beta,j} g_{\beta} a_{\beta}^{\dagger} a_{\beta} (b_{j}^{\dagger} + b_{j}) \nonumber\\
                 &\ -\sum_{\beta'>\beta,j} g_{\beta' \beta} (a_{\beta}^{\dagger} a_{\beta'} + a_{\beta} a_{\beta'}^{\dagger}) (b_{j}^{\dagger} + b_{j}),\\
    H_\mathrm{d} &= i \Omega \cos\theta (a_{-}^{\dagger} - a_{-}) - i \Omega \sin\theta(a_{+}^{\dagger} - a_{+}),
\end{align}
\end{subequations}
where $\beta \in {-,+,D}$ runs over the hybrid modes, and
\begin{subequations}
\begin{align}
  g_{-} &= \sin^{2}\theta \frac{g_{p}}{2}, & g_{+-} &= \sin\theta \cos\theta \frac{g_{p}}{2},\\
  g_{+} &= \cos^{2}\theta \frac{g_{p}}{2}, & g_{D-} &= \sin\theta \frac{g_{p}}{2},\\
  g_{D} &= \frac{g_{p}}{2},                & g_{D+} &= \cos\theta \frac{g_{p}}{2}.
\end{align}
\end{subequations}

\begin{figure}
  \includegraphics[width=\linewidth]{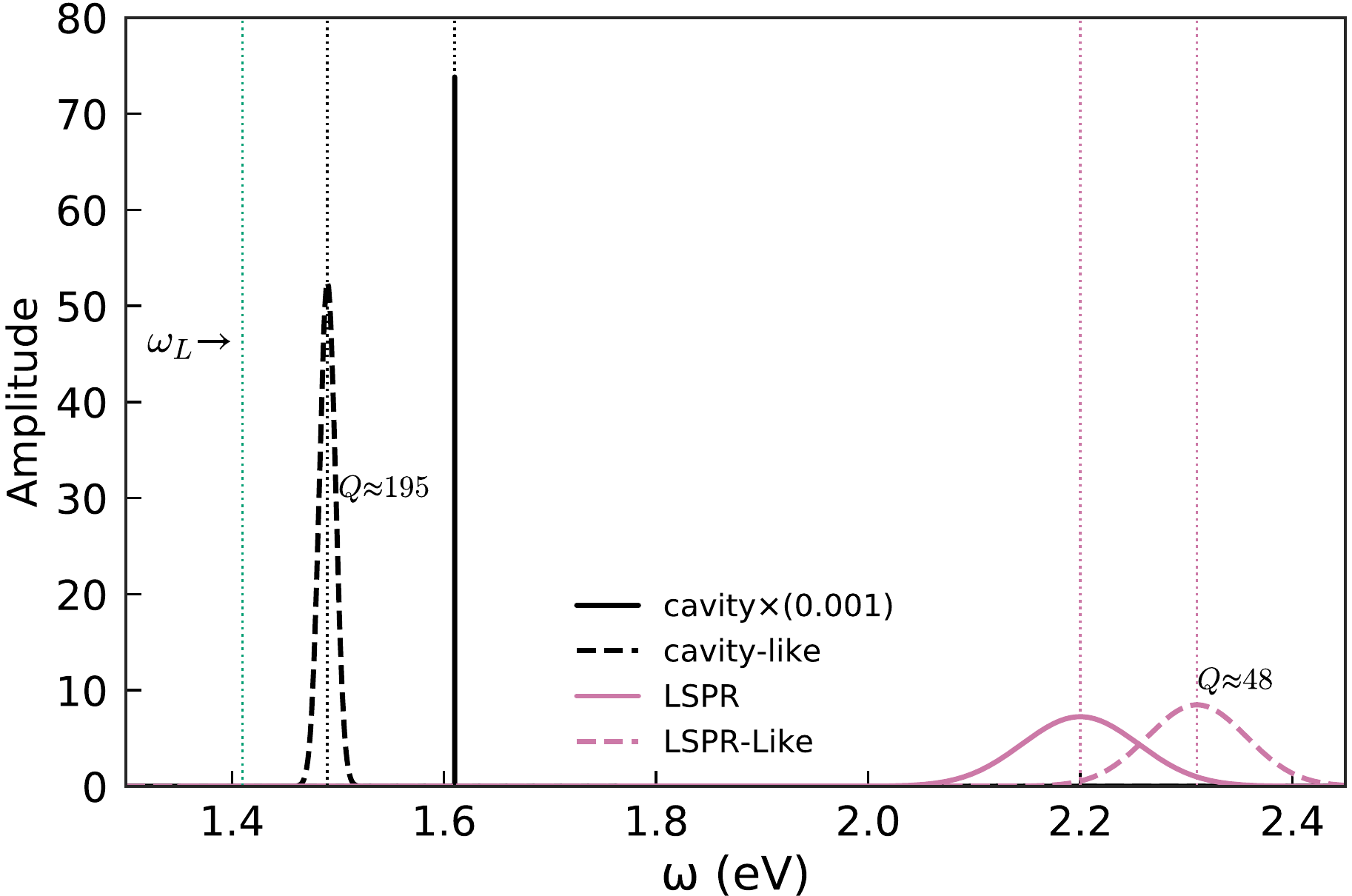}
  \caption{Comparison between the original (solid lines) and hybrid (dashed
  lines) modes of the system. Each mode is represented by a Lorentzian at
  frequency $\omega_\beta$ with linewidth $\kappa_\beta$ and amplitude
  $g_\beta$. The driving laser frequency $\omega_l$ is indicated by a thin
  blue-green dotted line.}
  \label{fig:hybrid_comp}
\end{figure}

The dynamics is then described by the master equation
\begin{equation}
  \frac{\mathrm{d}\rho}{\mathrm{d}t} = -i [H,\rho] + \sum_\beta L_{a_\beta}[\rho] 
  + \sum_j L_{b_{j}}[\rho] + \bar{L}_{\pm}[\rho],
\end{equation}
where
\begin{subequations}
\begin{align}
  L_{a_{\beta}}[\rho] &= \kappa_\beta D_{a_{\beta}}[\rho],\\
  \bar{L}_{\pm}[\rho] &= \kappa_{\pm} (D_{a_+,a_-}[\rho] + D_{a_-,a_+}[\rho]),
\end{align}
\end{subequations}
where $D_{a,b}[\rho] = a\rho b^{\dagger} - \frac12 \{ b^{\dagger} a, \rho \}$ and
\begin{subequations}
\begin{align}
  \kappa_{D} &= \kappa_{p},\\
  \kappa_{-} &= \cos^{2}\theta\kappa_{c} + \sin^{2}\theta\kappa_{p},\\
  \kappa_{+} &= \cos^{2}\theta\kappa_{p} + \sin^{2}\theta\kappa_{c},\\
  \kappa_{\pm} &= (\kappa_{p} - \kappa_{c}) \sin\theta \cos\theta.
\end{align}
\end{subequations}
The properties of the hybrid modes are shown schematically in
\autoref{fig:hybrid_comp}, where each mode is represented by a Lorentzian at
frequency $\omega_\beta = \delta_\beta + \omega_l$, with linewidth
$\kappa_\beta$ and amplitude $g_\beta$. The driving laser is indicates as a
dotted blue-green line at $\omega_l = 1.41$~eV. For the parameters considered,
the cavity-like mode $a_-$ thus clearly gives the most significant contribution
since it is close to resonance with the external laser and has a strong
optomechanical interaction. The simplified effective Hamiltonian obtained by
only taking into account the hybrid cavity-like mode is given by
\begin{multline}
  H_{\mathrm{hc}} \approx \delta_{-}a_{-}^{\dagger}a_{-} + \omega_{1}b_{1}^{\dagger}b_{1} + \omega_{2}b_{2}^{\dagger}b_{2} + i\Omega \cos\theta (a_{-}^{\dagger} - a_{-})\\
  - g_{-} a_{-}^{\dagger}a_{-} (b_{1}^{\dagger} + b_{1} + b_{2}^{\dagger} + b_{2}).
\end{multline}
The results obtained with this Hamiltonian are shown in the main text as the
``cavity-like'' approximation.

\bibliography{references}

\begin{thebibliography}{45}%
\makeatletter
\providecommand \@ifxundefined [1]{%
 \@ifx{#1\undefined}
}%
\providecommand \@ifnum [1]{%
 \ifnum #1\expandafter \@firstoftwo
 \else \expandafter \@secondoftwo
 \fi
}%
\providecommand \@ifx [1]{%
 \ifx #1\expandafter \@firstoftwo
 \else \expandafter \@secondoftwo
 \fi
}%
\providecommand \natexlab [1]{#1}%
\providecommand \enquote  [1]{``#1''}%
\providecommand \bibnamefont  [1]{#1}%
\providecommand \bibfnamefont [1]{#1}%
\providecommand \citenamefont [1]{#1}%
\providecommand \href@noop [0]{\@secondoftwo}%
\providecommand \href [0]{\begingroup \@sanitize@url \@href}%
\providecommand \@href[1]{\@@startlink{#1}\@@href}%
\providecommand \@@href[1]{\endgroup#1\@@endlink}%
\providecommand \@sanitize@url [0]{\catcode `\\12\catcode `\$12\catcode
  `\&12\catcode `\#12\catcode `\^12\catcode `\_12\catcode `\%12\relax}%
\providecommand \@@startlink[1]{}%
\providecommand \@@endlink[0]{}%
\providecommand \url  [0]{\begingroup\@sanitize@url \@url }%
\providecommand \@url [1]{\endgroup\@href {#1}{\urlprefix }}%
\providecommand \urlprefix  [0]{URL }%
\providecommand \Eprint [0]{\href }%
\providecommand \doibase [0]{https://doi.org/}%
\providecommand \selectlanguage [0]{\@gobble}%
\providecommand \bibinfo  [0]{\@secondoftwo}%
\providecommand \bibfield  [0]{\@secondoftwo}%
\providecommand \translation [1]{[#1]}%
\providecommand \BibitemOpen [0]{}%
\providecommand \bibitemStop [0]{}%
\providecommand \bibitemNoStop [0]{.\EOS\space}%
\providecommand \EOS [0]{\spacefactor3000\relax}%
\providecommand \BibitemShut  [1]{\csname bibitem#1\endcsname}%
\let\auto@bib@innerbib\@empty
\bibitem [{\citenamefont {Nagali}\ \emph {et~al.}(2009)\citenamefont {Nagali},
  \citenamefont {Sciarrino}, \citenamefont {De~Martini}, \citenamefont
  {Marrucci}, \citenamefont {Piccirillo}, \citenamefont {Karimi},\ and\
  \citenamefont {Santamato}}]{Nagali2009}%
  \BibitemOpen
  \bibfield  {author} {\bibinfo {author} {\bibfnamefont {E.}~\bibnamefont
  {Nagali}}, \bibinfo {author} {\bibfnamefont {F.}~\bibnamefont {Sciarrino}},
  \bibinfo {author} {\bibfnamefont {F.}~\bibnamefont {De~Martini}}, \bibinfo
  {author} {\bibfnamefont {L.}~\bibnamefont {Marrucci}}, \bibinfo {author}
  {\bibfnamefont {B.}~\bibnamefont {Piccirillo}}, \bibinfo {author}
  {\bibfnamefont {E.}~\bibnamefont {Karimi}},\ and\ \bibinfo {author}
  {\bibfnamefont {E.}~\bibnamefont {Santamato}},\ }\bibfield  {title} {\bibinfo
  {title} {{Quantum {{Information Transfer}} from {{Spin}} to {{Orbital Angular
  Momentum}} of {{Photons}}}},\ }\href
  {https://doi.org/10.1103/PhysRevLett.103.013601} {\bibfield  {journal}
  {\bibinfo  {journal} {Phys. Rev. Lett.}\ }\textbf {\bibinfo {volume} {103}},\
  \bibinfo {pages} {013601} (\bibinfo {year} {2009})}\BibitemShut {NoStop}%
\bibitem [{\citenamefont {Northup}\ and\ \citenamefont
  {Blatt}(2014)}]{Northup2014}%
  \BibitemOpen
  \bibfield  {author} {\bibinfo {author} {\bibfnamefont {T.~E.}\ \bibnamefont
  {Northup}}\ and\ \bibinfo {author} {\bibfnamefont {R.}~\bibnamefont
  {Blatt}},\ }\bibfield  {title} {\bibinfo {title} {{Quantum Information
  Transfer Using Photons}},\ }\href {https://doi.org/10.1038/nphoton.2014.53}
  {\bibfield  {journal} {\bibinfo  {journal} {Nat. Photonics}\ }\textbf
  {\bibinfo {volume} {8}},\ \bibinfo {pages} {356} (\bibinfo {year}
  {2014})}\BibitemShut {NoStop}%
\bibitem [{\citenamefont {Nalbach}\ \emph {et~al.}(2010)\citenamefont
  {Nalbach}, \citenamefont {Eckel},\ and\ \citenamefont
  {Thorwart}}]{Nalbach2010}%
  \BibitemOpen
  \bibfield  {author} {\bibinfo {author} {\bibfnamefont {P.}~\bibnamefont
  {Nalbach}}, \bibinfo {author} {\bibfnamefont {J.}~\bibnamefont {Eckel}},\
  and\ \bibinfo {author} {\bibfnamefont {M.}~\bibnamefont {Thorwart}},\
  }\bibfield  {title} {\bibinfo {title} {{Quantum Coherent Biomolecular Energy
  Transfer with Spatially Correlated Fluctuations}},\ }\href
  {https://doi.org/10.1088/1367-2630/12/6/065043} {\bibfield  {journal}
  {\bibinfo  {journal} {New J. Phys.}\ }\textbf {\bibinfo {volume} {12}},\
  \bibinfo {pages} {065043} (\bibinfo {year} {2010})}\BibitemShut {NoStop}%
\bibitem [{\citenamefont {Dubi}\ and\ \citenamefont
  {Di~Ventra}(2011)}]{Dubi2011}%
  \BibitemOpen
  \bibfield  {author} {\bibinfo {author} {\bibfnamefont {Y.}~\bibnamefont
  {Dubi}}\ and\ \bibinfo {author} {\bibfnamefont {M.}~\bibnamefont
  {Di~Ventra}},\ }\bibfield  {title} {\bibinfo {title} {{Colloquium: {{Heat}}
  Flow and Thermoelectricity in Atomic and Molecular Junctions}},\ }\href
  {https://doi.org/10.1103/RevModPhys.83.131} {\bibfield  {journal} {\bibinfo
  {journal} {Rev. Mod. Phys.}\ }\textbf {\bibinfo {volume} {83}},\ \bibinfo
  {pages} {131} (\bibinfo {year} {2011})}\BibitemShut {NoStop}%
\bibitem [{\citenamefont {Katz}\ and\ \citenamefont
  {Kosloff}(2016)}]{Katz2016}%
  \BibitemOpen
  \bibfield  {author} {\bibinfo {author} {\bibfnamefont {G.}~\bibnamefont
  {Katz}}\ and\ \bibinfo {author} {\bibfnamefont {R.}~\bibnamefont {Kosloff}},\
  }\bibfield  {title} {\bibinfo {title} {{Quantum {{Thermodynamics}} in
  {{Strong Coupling}}: {{Heat Transport}} and {{Refrigeration}}}},\ }\href
  {https://doi.org/10.3390/e18050186} {\bibfield  {journal} {\bibinfo
  {journal} {Entropy}\ }\textbf {\bibinfo {volume} {18}},\ \bibinfo {pages}
  {186} (\bibinfo {year} {2016})}\BibitemShut {NoStop}%
\bibitem [{\citenamefont {Lee}\ \emph {et~al.}(2007)\citenamefont {Lee},
  \citenamefont {Cheng},\ and\ \citenamefont {Fleming}}]{Lee2007}%
  \BibitemOpen
  \bibfield  {author} {\bibinfo {author} {\bibfnamefont {H.}~\bibnamefont
  {Lee}}, \bibinfo {author} {\bibfnamefont {Y.-C.}\ \bibnamefont {Cheng}},\
  and\ \bibinfo {author} {\bibfnamefont {G.~R.}\ \bibnamefont {Fleming}},\
  }\bibfield  {title} {\bibinfo {title} {{Coherence Dynamics in Photosynthesis:
  Protein Protection of Excitonic Coherence}},\ }\href
  {https://doi.org/10.1126/science.1142188} {\bibfield  {journal} {\bibinfo
  {journal} {Science}\ }\textbf {\bibinfo {volume} {316}},\ \bibinfo {pages}
  {1462} (\bibinfo {year} {2007})}\BibitemShut {NoStop}%
\bibitem [{\citenamefont {Scholes}\ \emph {et~al.}(2011)\citenamefont
  {Scholes}, \citenamefont {Fleming}, \citenamefont {{Olaya-Castro}},\ and\
  \citenamefont {{van Grondelle}}}]{Scholes2011}%
  \BibitemOpen
  \bibfield  {author} {\bibinfo {author} {\bibfnamefont {G.~D.}\ \bibnamefont
  {Scholes}}, \bibinfo {author} {\bibfnamefont {G.~R.}\ \bibnamefont
  {Fleming}}, \bibinfo {author} {\bibfnamefont {A.}~\bibnamefont
  {{Olaya-Castro}}},\ and\ \bibinfo {author} {\bibfnamefont {R.}~\bibnamefont
  {{van Grondelle}}},\ }\bibfield  {title} {\bibinfo {title} {{Lessons from
  Nature about Solar Light Harvesting}},\ }\href
  {https://doi.org/10.1038/nchem.1145} {\bibfield  {journal} {\bibinfo
  {journal} {Nat. Chem.}\ }\textbf {\bibinfo {volume} {3}},\ \bibinfo {pages}
  {763} (\bibinfo {year} {2011})}\BibitemShut {NoStop}%
\bibitem [{\citenamefont {High}\ \emph {et~al.}(2008)\citenamefont {High},
  \citenamefont {Novitskaya}, \citenamefont {Butov}, \citenamefont {Hanson},\
  and\ \citenamefont {Gossard}}]{High2008}%
  \BibitemOpen
  \bibfield  {author} {\bibinfo {author} {\bibfnamefont {A.~A.}\ \bibnamefont
  {High}}, \bibinfo {author} {\bibfnamefont {E.~E.}\ \bibnamefont
  {Novitskaya}}, \bibinfo {author} {\bibfnamefont {L.~V.}\ \bibnamefont
  {Butov}}, \bibinfo {author} {\bibfnamefont {M.}~\bibnamefont {Hanson}},\ and\
  \bibinfo {author} {\bibfnamefont {A.~C.}\ \bibnamefont {Gossard}},\
  }\bibfield  {title} {\bibinfo {title} {{Control of Exciton Fluxes in an
  Excitonic Integrated Circuit}},\ }\href
  {https://doi.org/10.1126/science.1157845} {\bibfield  {journal} {\bibinfo
  {journal} {Science}\ }\textbf {\bibinfo {volume} {321}},\ \bibinfo {pages}
  {229} (\bibinfo {year} {2008})}\BibitemShut {NoStop}%
\bibitem [{\citenamefont {Menke}\ \emph {et~al.}(2012)\citenamefont {Menke},
  \citenamefont {Luhman},\ and\ \citenamefont {Holmes}}]{Menke2013}%
  \BibitemOpen
  \bibfield  {author} {\bibinfo {author} {\bibfnamefont {S.~M.}\ \bibnamefont
  {Menke}}, \bibinfo {author} {\bibfnamefont {W.~A.}\ \bibnamefont {Luhman}},\
  and\ \bibinfo {author} {\bibfnamefont {R.~J.}\ \bibnamefont {Holmes}},\
  }\bibfield  {title} {\bibinfo {title} {{Tailored Exciton Diffusion in Organic
  Photovoltaic Cells for Enhanced Power Conversion Efficiency}},\ }\href
  {https://doi.org/10.1038/nmat3467} {\bibfield  {journal} {\bibinfo  {journal}
  {Nat. Mater.}\ }\textbf {\bibinfo {volume} {12}},\ \bibinfo {pages} {152}
  (\bibinfo {year} {2012})}\BibitemShut {NoStop}%
\bibitem [{\citenamefont {Gerry}\ and\ \citenamefont
  {Knight}(2004)}]{Gerry2004}%
  \BibitemOpen
  \bibfield  {author} {\bibinfo {author} {\bibfnamefont {C.}~\bibnamefont
  {Gerry}}\ and\ \bibinfo {author} {\bibfnamefont {P.}~\bibnamefont {Knight}},\
  }\href {https://doi.org/10.1017/CBO9780511791239} {\emph {\bibinfo {title}
  {{Introductory {{Quantum Optics}}}}}}\ (\bibinfo  {publisher} {{Cambridge
  University Press}},\ \bibinfo {address} {{Cambridge}},\ \bibinfo {year}
  {2004})\BibitemShut {NoStop}%
\bibitem [{\citenamefont {Messina}\ \emph {et~al.}(2012)\citenamefont
  {Messina}, \citenamefont {Antezza},\ and\ \citenamefont
  {{Ben-Abdallah}}}]{Messina2012}%
  \BibitemOpen
  \bibfield  {author} {\bibinfo {author} {\bibfnamefont {R.}~\bibnamefont
  {Messina}}, \bibinfo {author} {\bibfnamefont {M.}~\bibnamefont {Antezza}},\
  and\ \bibinfo {author} {\bibfnamefont {P.}~\bibnamefont {{Ben-Abdallah}}},\
  }\bibfield  {title} {\bibinfo {title} {{Three-{{Body Amplification}} of
  {{Photon Heat Tunneling}}}},\ }\href
  {https://doi.org/10.1103/PhysRevLett.109.244302} {\bibfield  {journal}
  {\bibinfo  {journal} {Phys. Rev. Lett.}\ }\textbf {\bibinfo {volume} {109}},\
  \bibinfo {pages} {244302} (\bibinfo {year} {2012})}\BibitemShut {NoStop}%
\bibitem [{\citenamefont {Feist}\ and\ \citenamefont
  {{Garcia-Vidal}}(2015)}]{Feist2015}%
  \BibitemOpen
  \bibfield  {author} {\bibinfo {author} {\bibfnamefont {J.}~\bibnamefont
  {Feist}}\ and\ \bibinfo {author} {\bibfnamefont {F.~J.}\ \bibnamefont
  {{Garcia-Vidal}}},\ }\bibfield  {title} {\bibinfo {title} {{Extraordinary
  {{Exciton Conductance Induced}} by {{Strong Coupling}}}},\ }\href
  {https://doi.org/10.1103/PhysRevLett.114.196402} {\bibfield  {journal}
  {\bibinfo  {journal} {Phys. Rev. Lett.}\ }\textbf {\bibinfo {volume} {114}},\
  \bibinfo {pages} {196402} (\bibinfo {year} {2015})}\BibitemShut {NoStop}%
\bibitem [{\citenamefont {Schachenmayer}\ \emph {et~al.}(2015)\citenamefont
  {Schachenmayer}, \citenamefont {Genes}, \citenamefont {Tignone},\ and\
  \citenamefont {Pupillo}}]{Schachenmayer2015}%
  \BibitemOpen
  \bibfield  {author} {\bibinfo {author} {\bibfnamefont {J.}~\bibnamefont
  {Schachenmayer}}, \bibinfo {author} {\bibfnamefont {C.}~\bibnamefont
  {Genes}}, \bibinfo {author} {\bibfnamefont {E.}~\bibnamefont {Tignone}},\
  and\ \bibinfo {author} {\bibfnamefont {G.}~\bibnamefont {Pupillo}},\
  }\bibfield  {title} {\bibinfo {title} {{Cavity-{{Enhanced Transport}} of
  {{Excitons}}}},\ }\href {https://doi.org/10.1103/PhysRevLett.114.196403}
  {\bibfield  {journal} {\bibinfo  {journal} {Phys. Rev. Lett.}\ }\textbf
  {\bibinfo {volume} {114}},\ \bibinfo {pages} {196403} (\bibinfo {year}
  {2015})}\BibitemShut {NoStop}%
\bibitem [{\citenamefont {Kim}\ \emph {et~al.}(2015)\citenamefont {Kim},
  \citenamefont {Sim}, \citenamefont {Yoon}, \citenamefont {Gong},
  \citenamefont {Ahn}, \citenamefont {Cho},\ and\ \citenamefont
  {Lee}}]{Kim2015}%
  \BibitemOpen
  \bibfield  {author} {\bibinfo {author} {\bibfnamefont {M.-K.}\ \bibnamefont
  {Kim}}, \bibinfo {author} {\bibfnamefont {H.}~\bibnamefont {Sim}}, \bibinfo
  {author} {\bibfnamefont {S.~J.}\ \bibnamefont {Yoon}}, \bibinfo {author}
  {\bibfnamefont {S.-H.}\ \bibnamefont {Gong}}, \bibinfo {author}
  {\bibfnamefont {C.~W.}\ \bibnamefont {Ahn}}, \bibinfo {author} {\bibfnamefont
  {Y.-H.}\ \bibnamefont {Cho}},\ and\ \bibinfo {author} {\bibfnamefont {Y.-H.}\
  \bibnamefont {Lee}},\ }\bibfield  {title} {\bibinfo {title} {{Squeezing
  {{Photons}} into a {{Point}}-{{Like Space}}}},\ }\href
  {https://doi.org/10.1021/acs.nanolett.5b01204} {\bibfield  {journal}
  {\bibinfo  {journal} {Nano Lett.}\ }\textbf {\bibinfo {volume} {15}},\
  \bibinfo {pages} {4102} (\bibinfo {year} {2015})}\BibitemShut {NoStop}%
\bibitem [{\citenamefont {Chikkaraddy}\ \emph {et~al.}(2016)\citenamefont
  {Chikkaraddy}, \citenamefont {{de Nijs}}, \citenamefont {Benz}, \citenamefont
  {Barrow}, \citenamefont {Scherman}, \citenamefont {Rosta}, \citenamefont
  {Demetriadou}, \citenamefont {Fox}, \citenamefont {Hess},\ and\ \citenamefont
  {Baumberg}}]{Chikkaraddy2016}%
  \BibitemOpen
  \bibfield  {author} {\bibinfo {author} {\bibfnamefont {R.}~\bibnamefont
  {Chikkaraddy}}, \bibinfo {author} {\bibfnamefont {B.}~\bibnamefont {{de
  Nijs}}}, \bibinfo {author} {\bibfnamefont {F.}~\bibnamefont {Benz}}, \bibinfo
  {author} {\bibfnamefont {S.~J.}\ \bibnamefont {Barrow}}, \bibinfo {author}
  {\bibfnamefont {O.~A.}\ \bibnamefont {Scherman}}, \bibinfo {author}
  {\bibfnamefont {E.}~\bibnamefont {Rosta}}, \bibinfo {author} {\bibfnamefont
  {A.}~\bibnamefont {Demetriadou}}, \bibinfo {author} {\bibfnamefont
  {P.}~\bibnamefont {Fox}}, \bibinfo {author} {\bibfnamefont {O.}~\bibnamefont
  {Hess}},\ and\ \bibinfo {author} {\bibfnamefont {J.~J.}\ \bibnamefont
  {Baumberg}},\ }\bibfield  {title} {\bibinfo {title} {{Single-Molecule Strong
  Coupling at Room Temperature in Plasmonic Nanocavities}},\ }\href
  {https://doi.org/10.1038/nature17974} {\bibfield  {journal} {\bibinfo
  {journal} {Nature}\ }\textbf {\bibinfo {volume} {535}},\ \bibinfo {pages}
  {127} (\bibinfo {year} {2016})}\BibitemShut {NoStop}%
\bibitem [{\citenamefont {Ojambati}\ \emph {et~al.}(2019)\citenamefont
  {Ojambati}, \citenamefont {Chikkaraddy}, \citenamefont {Deacon},
  \citenamefont {Horton}, \citenamefont {Kos}, \citenamefont {Turek},
  \citenamefont {Keyser},\ and\ \citenamefont {Baumberg}}]{Ojambati2019}%
  \BibitemOpen
  \bibfield  {author} {\bibinfo {author} {\bibfnamefont {O.~S.}\ \bibnamefont
  {Ojambati}}, \bibinfo {author} {\bibfnamefont {R.}~\bibnamefont
  {Chikkaraddy}}, \bibinfo {author} {\bibfnamefont {W.~D.}\ \bibnamefont
  {Deacon}}, \bibinfo {author} {\bibfnamefont {M.}~\bibnamefont {Horton}},
  \bibinfo {author} {\bibfnamefont {D.}~\bibnamefont {Kos}}, \bibinfo {author}
  {\bibfnamefont {V.~A.}\ \bibnamefont {Turek}}, \bibinfo {author}
  {\bibfnamefont {U.~F.}\ \bibnamefont {Keyser}},\ and\ \bibinfo {author}
  {\bibfnamefont {J.~J.}\ \bibnamefont {Baumberg}},\ }\bibfield  {title}
  {\bibinfo {title} {{Quantum Electrodynamics at Room Temperature Coupling a
  Single Vibrating Molecule with a Plasmonic Nanocavity}},\ }\href
  {https://doi.org/10.1038/s41467-019-08611-5} {\bibfield  {journal} {\bibinfo
  {journal} {Nat. Commun.}\ }\textbf {\bibinfo {volume} {10}},\ \bibinfo
  {pages} {1049} (\bibinfo {year} {2019})}\BibitemShut {NoStop}%
\bibitem [{\citenamefont {{Fern{\'a}ndez-Dom{\'i}nguez}}\ \emph
  {et~al.}(2017)\citenamefont {{Fern{\'a}ndez-Dom{\'i}nguez}}, \citenamefont
  {{Garc{\'i}a-Vidal}},\ and\ \citenamefont
  {{Mart{\'i}n-Moreno}}}]{Fernandez-Dominguez2017}%
  \BibitemOpen
  \bibfield  {author} {\bibinfo {author} {\bibfnamefont {A.~I.}\ \bibnamefont
  {{Fern{\'a}ndez-Dom{\'i}nguez}}}, \bibinfo {author} {\bibfnamefont {F.~J.}\
  \bibnamefont {{Garc{\'i}a-Vidal}}},\ and\ \bibinfo {author} {\bibfnamefont
  {L.}~\bibnamefont {{Mart{\'i}n-Moreno}}},\ }\bibfield  {title} {\bibinfo
  {title} {{Unrelenting Plasmons}},\ }\href
  {https://doi.org/10.1038/nphoton.2016.258} {\bibfield  {journal} {\bibinfo
  {journal} {Nat. Photonics}\ }\textbf {\bibinfo {volume} {11}},\ \bibinfo
  {pages} {8} (\bibinfo {year} {2017})}\BibitemShut {NoStop}%
\bibitem [{\citenamefont {Hoang}\ \emph {et~al.}(2016)\citenamefont {Hoang},
  \citenamefont {Akselrod},\ and\ \citenamefont {Mikkelsen}}]{Hoang2016}%
  \BibitemOpen
  \bibfield  {author} {\bibinfo {author} {\bibfnamefont {T.~B.}\ \bibnamefont
  {Hoang}}, \bibinfo {author} {\bibfnamefont {G.~M.}\ \bibnamefont
  {Akselrod}},\ and\ \bibinfo {author} {\bibfnamefont {M.~H.}\ \bibnamefont
  {Mikkelsen}},\ }\bibfield  {title} {\bibinfo {title} {{Ultrafast
  {{Room}}-{{Temperature Single Photon Emission}} from {{Quantum Dots Coupled}}
  to {{Plasmonic Nanocavities}}}},\ }\href
  {https://doi.org/10.1021/acs.nanolett.5b03724} {\bibfield  {journal}
  {\bibinfo  {journal} {Nano Lett.}\ }\textbf {\bibinfo {volume} {16}},\
  \bibinfo {pages} {270} (\bibinfo {year} {2016})}\BibitemShut {NoStop}%
\bibitem [{\citenamefont {Straubel}\ \emph {et~al.}(2017)\citenamefont
  {Straubel}, \citenamefont {Sarniak}, \citenamefont {Rockstuhl},\ and\
  \citenamefont {S{\l}owik}}]{Straubel2017}%
  \BibitemOpen
  \bibfield  {author} {\bibinfo {author} {\bibfnamefont {J.}~\bibnamefont
  {Straubel}}, \bibinfo {author} {\bibfnamefont {R.}~\bibnamefont {Sarniak}},
  \bibinfo {author} {\bibfnamefont {C.}~\bibnamefont {Rockstuhl}},\ and\
  \bibinfo {author} {\bibfnamefont {K.}~\bibnamefont {S{\l}owik}},\ }\bibfield
  {title} {\bibinfo {title} {{Entangled Light from Bimodal Optical
  Nanoantennas}},\ }\href {https://doi.org/10.1103/PhysRevB.95.085421}
  {\bibfield  {journal} {\bibinfo  {journal} {Phys. Rev. B}\ }\textbf {\bibinfo
  {volume} {95}},\ \bibinfo {pages} {085421} (\bibinfo {year}
  {2017})}\BibitemShut {NoStop}%
\bibitem [{\citenamefont {Kneipp}\ \emph {et~al.}(2006)\citenamefont {Kneipp},
  \citenamefont {Moskovits},\ and\ \citenamefont {Kneipp}}]{Kneipp2006}%
  \BibitemOpen
  \bibinfo {editor} {\bibfnamefont {K.}~\bibnamefont {Kneipp}}, \bibinfo
  {editor} {\bibfnamefont {M.}~\bibnamefont {Moskovits}},\ and\ \bibinfo
  {editor} {\bibfnamefont {H.}~\bibnamefont {Kneipp}},\ eds.,\ \href@noop {}
  {\emph {\bibinfo {title} {{Surface-{{Enhanced Raman Scattering}}: {{Physics}}
  and {{Applications}} ({{Topics}} in {{Applied Physics}})}}}},\ \bibinfo
  {edition} {1st}\ ed.\ (\bibinfo  {publisher} {{Springer}},\ \bibinfo {year}
  {2006})\BibitemShut {NoStop}%
\bibitem [{\citenamefont {Schl{\"u}cker}(2014)}]{Schlucker2014}%
  \BibitemOpen
  \bibfield  {author} {\bibinfo {author} {\bibfnamefont {S.}~\bibnamefont
  {Schl{\"u}cker}},\ }\bibfield  {title} {\bibinfo {title} {{Surface-{{Enhanced
  Raman Spectroscopy}}: {{Concepts}} and {{Chemical Applications}}}},\ }\href
  {https://doi.org/10.1002/anie.201205748} {\bibfield  {journal} {\bibinfo
  {journal} {Angew. Chem. Int. Ed.}\ }\textbf {\bibinfo {volume} {53}},\
  \bibinfo {pages} {4756} (\bibinfo {year} {2014})}\BibitemShut {NoStop}%
\bibitem [{\citenamefont {Roelli}\ \emph {et~al.}(2016)\citenamefont {Roelli},
  \citenamefont {Galland}, \citenamefont {Piro},\ and\ \citenamefont
  {Kippenberg}}]{Roelli2016}%
  \BibitemOpen
  \bibfield  {author} {\bibinfo {author} {\bibfnamefont {P.}~\bibnamefont
  {Roelli}}, \bibinfo {author} {\bibfnamefont {C.}~\bibnamefont {Galland}},
  \bibinfo {author} {\bibfnamefont {N.}~\bibnamefont {Piro}},\ and\ \bibinfo
  {author} {\bibfnamefont {T.~J.}\ \bibnamefont {Kippenberg}},\ }\bibfield
  {title} {\bibinfo {title} {{Molecular Cavity Optomechanics as a Theory of
  Plasmon-Enhanced {{Raman}} Scattering}},\ }\href
  {https://doi.org/10.1038/nnano.2015.264} {\bibfield  {journal} {\bibinfo
  {journal} {Nat. Nanotechnol.}\ }\textbf {\bibinfo {volume} {11}},\ \bibinfo
  {pages} {164} (\bibinfo {year} {2016})}\BibitemShut {NoStop}%
\bibitem [{\citenamefont {Schmidt}\ \emph {et~al.}(2016)\citenamefont
  {Schmidt}, \citenamefont {Esteban}, \citenamefont {{Gonz{\'a}lez-Tudela}},
  \citenamefont {Giedke},\ and\ \citenamefont {Aizpurua}}]{Schmidt2016Quantum}%
  \BibitemOpen
  \bibfield  {author} {\bibinfo {author} {\bibfnamefont {M.~K.}\ \bibnamefont
  {Schmidt}}, \bibinfo {author} {\bibfnamefont {R.}~\bibnamefont {Esteban}},
  \bibinfo {author} {\bibfnamefont {A.}~\bibnamefont {{Gonz{\'a}lez-Tudela}}},
  \bibinfo {author} {\bibfnamefont {G.}~\bibnamefont {Giedke}},\ and\ \bibinfo
  {author} {\bibfnamefont {J.}~\bibnamefont {Aizpurua}},\ }\bibfield  {title}
  {\bibinfo {title} {{Quantum {{Mechanical Description}} of {{Raman
  Scattering}} from {{Molecules}} in {{Plasmonic Cavities}}}},\ }\href
  {https://doi.org/10.1021/acsnano.6b02484} {\bibfield  {journal} {\bibinfo
  {journal} {ACS Nano}\ }\textbf {\bibinfo {volume} {10}},\ \bibinfo {pages}
  {6291} (\bibinfo {year} {2016})}\BibitemShut {NoStop}%
\bibitem [{\citenamefont {Ashrafi}\ \emph {et~al.}(2019)\citenamefont
  {Ashrafi}, \citenamefont {Malekfar}, \citenamefont {Bahrampour},\ and\
  \citenamefont {Feist}}]{Ashrafi2019}%
  \BibitemOpen
  \bibfield  {author} {\bibinfo {author} {\bibfnamefont {S.~M.}\ \bibnamefont
  {Ashrafi}}, \bibinfo {author} {\bibfnamefont {R.}~\bibnamefont {Malekfar}},
  \bibinfo {author} {\bibfnamefont {A.~R.}\ \bibnamefont {Bahrampour}},\ and\
  \bibinfo {author} {\bibfnamefont {J.}~\bibnamefont {Feist}},\ }\bibfield
  {title} {\bibinfo {title} {{Optomechanical Heat Transfer between Molecules in
  a Nanoplasmonic Cavity}},\ }\href
  {https://doi.org/10.1103/PhysRevA.100.013826} {\bibfield  {journal} {\bibinfo
   {journal} {Phys. Rev. A}\ }\textbf {\bibinfo {volume} {100}},\ \bibinfo
  {pages} {013826} (\bibinfo {year} {2019})}\BibitemShut {NoStop}%
\bibitem [{\citenamefont {Xiao}\ \emph {et~al.}(2012)\citenamefont {Xiao},
  \citenamefont {Liu}, \citenamefont {Li}, \citenamefont {Chen}, \citenamefont
  {Li},\ and\ \citenamefont {Gong}}]{Xiao2012}%
  \BibitemOpen
  \bibfield  {author} {\bibinfo {author} {\bibfnamefont {Y.-F.}\ \bibnamefont
  {Xiao}}, \bibinfo {author} {\bibfnamefont {Y.-C.}\ \bibnamefont {Liu}},
  \bibinfo {author} {\bibfnamefont {B.-B.}\ \bibnamefont {Li}}, \bibinfo
  {author} {\bibfnamefont {Y.-L.}\ \bibnamefont {Chen}}, \bibinfo {author}
  {\bibfnamefont {Y.}~\bibnamefont {Li}},\ and\ \bibinfo {author}
  {\bibfnamefont {Q.}~\bibnamefont {Gong}},\ }\bibfield  {title} {\bibinfo
  {title} {{Strongly Enhanced Light-Matter Interaction in a Hybrid
  Photonic-Plasmonic Resonator}},\ }\href
  {https://doi.org/10.1103/PhysRevA.85.031805} {\bibfield  {journal} {\bibinfo
  {journal} {Phys. Rev. A}\ }\textbf {\bibinfo {volume} {85}},\ \bibinfo
  {pages} {031805} (\bibinfo {year} {2012})}\BibitemShut {NoStop}%
\bibitem [{\citenamefont {Conteduca}\ \emph {et~al.}(2017)\citenamefont
  {Conteduca}, \citenamefont {Reardon}, \citenamefont {Scullion}, \citenamefont
  {Dell'Olio}, \citenamefont {Armenise}, \citenamefont {Krauss},\ and\
  \citenamefont {Ciminelli}}]{Conteduca2017}%
  \BibitemOpen
  \bibfield  {author} {\bibinfo {author} {\bibfnamefont {D.}~\bibnamefont
  {Conteduca}}, \bibinfo {author} {\bibfnamefont {C.}~\bibnamefont {Reardon}},
  \bibinfo {author} {\bibfnamefont {M.~G.}\ \bibnamefont {Scullion}}, \bibinfo
  {author} {\bibfnamefont {F.}~\bibnamefont {Dell'Olio}}, \bibinfo {author}
  {\bibfnamefont {M.~N.}\ \bibnamefont {Armenise}}, \bibinfo {author}
  {\bibfnamefont {T.~F.}\ \bibnamefont {Krauss}},\ and\ \bibinfo {author}
  {\bibfnamefont {C.}~\bibnamefont {Ciminelli}},\ }\bibfield  {title} {\bibinfo
  {title} {{Ultra-High {{Q}}/{{V}} Hybrid Cavity for Strong Light-Matter
  Interaction}},\ }\href {https://doi.org/10.1063/1.4994056} {\bibfield
  {journal} {\bibinfo  {journal} {APL Photonics}\ }\textbf {\bibinfo {volume}
  {2}},\ \bibinfo {pages} {086101} (\bibinfo {year} {2017})}\BibitemShut
  {NoStop}%
\bibitem [{\citenamefont {Gurlek}\ \emph {et~al.}(2018)\citenamefont {Gurlek},
  \citenamefont {Sandoghdar},\ and\ \citenamefont
  {{Mart{\'i}n-Cano}}}]{Gurlek2018}%
  \BibitemOpen
  \bibfield  {author} {\bibinfo {author} {\bibfnamefont {B.}~\bibnamefont
  {Gurlek}}, \bibinfo {author} {\bibfnamefont {V.}~\bibnamefont {Sandoghdar}},\
  and\ \bibinfo {author} {\bibfnamefont {D.}~\bibnamefont
  {{Mart{\'i}n-Cano}}},\ }\bibfield  {title} {\bibinfo {title} {{Manipulation
  of {{Quenching}} in {{Nanoantenna}}\textendash{{Emitter Systems Enabled}} by
  {{External Detuned Cavities}}: {{A Path}} to {{Enhance
  Strong}}-{{Coupling}}}},\ }\href
  {https://doi.org/10.1021/acsphotonics.7b00953} {\bibfield  {journal}
  {\bibinfo  {journal} {ACS Photonics}\ }\textbf {\bibinfo {volume} {5}},\
  \bibinfo {pages} {456} (\bibinfo {year} {2018})}\BibitemShut {NoStop}%
\bibitem [{\citenamefont {Mossayebi}\ \emph {et~al.}(2016)\citenamefont
  {Mossayebi}, \citenamefont {Wright}, \citenamefont {Parini}, \citenamefont
  {Somekh}, \citenamefont {Bellanca},\ and\ \citenamefont
  {Larkins}}]{Mossayebi2016}%
  \BibitemOpen
  \bibfield  {author} {\bibinfo {author} {\bibfnamefont {M.}~\bibnamefont
  {Mossayebi}}, \bibinfo {author} {\bibfnamefont {A.~J.}\ \bibnamefont
  {Wright}}, \bibinfo {author} {\bibfnamefont {A.}~\bibnamefont {Parini}},
  \bibinfo {author} {\bibfnamefont {M.~G.}\ \bibnamefont {Somekh}}, \bibinfo
  {author} {\bibfnamefont {G.}~\bibnamefont {Bellanca}},\ and\ \bibinfo
  {author} {\bibfnamefont {E.~C.}\ \bibnamefont {Larkins}},\ }\bibfield
  {title} {\bibinfo {title} {{Investigating the Use of a Hybrid
  Plasmonic\textendash Photonic Nanoresonator for Optical Trapping Using
  Finite-Difference Time-Domain Method}},\ }\href
  {https://doi.org/10.1007/s11082-016-0539-5} {\bibfield  {journal} {\bibinfo
  {journal} {Opt Quant Electron}\ }\textbf {\bibinfo {volume} {48}},\ \bibinfo
  {pages} {275} (\bibinfo {year} {2016})}\BibitemShut {NoStop}%
\bibitem [{\citenamefont {Peyskens}\ \emph {et~al.}(2016)\citenamefont
  {Peyskens}, \citenamefont {Dhakal}, \citenamefont {Van~Dorpe}, \citenamefont
  {Le~Thomas},\ and\ \citenamefont {Baets}}]{Peyskens2016}%
  \BibitemOpen
  \bibfield  {author} {\bibinfo {author} {\bibfnamefont {F.}~\bibnamefont
  {Peyskens}}, \bibinfo {author} {\bibfnamefont {A.}~\bibnamefont {Dhakal}},
  \bibinfo {author} {\bibfnamefont {P.}~\bibnamefont {Van~Dorpe}}, \bibinfo
  {author} {\bibfnamefont {N.}~\bibnamefont {Le~Thomas}},\ and\ \bibinfo
  {author} {\bibfnamefont {R.}~\bibnamefont {Baets}},\ }\bibfield  {title}
  {\bibinfo {title} {{Surface {{Enhanced Raman Spectroscopy Using}} a {{Single
  Mode Nanophotonic}}-{{Plasmonic Platform}}}},\ }\href
  {https://doi.org/10.1021/acsphotonics.5b00487} {\bibfield  {journal}
  {\bibinfo  {journal} {ACS Photonics}\ }\textbf {\bibinfo {volume} {3}},\
  \bibinfo {pages} {102} (\bibinfo {year} {2016})}\BibitemShut {NoStop}%
\bibitem [{\citenamefont {De~Angelis}\ \emph {et~al.}(2008)\citenamefont
  {De~Angelis}, \citenamefont {Patrini}, \citenamefont {Das}, \citenamefont
  {Maksymov}, \citenamefont {Galli}, \citenamefont {Businaro}, \citenamefont
  {Andreani},\ and\ \citenamefont {Di~Fabrizio}}]{DeAngelis2008}%
  \BibitemOpen
  \bibfield  {author} {\bibinfo {author} {\bibfnamefont {F.}~\bibnamefont
  {De~Angelis}}, \bibinfo {author} {\bibfnamefont {M.}~\bibnamefont {Patrini}},
  \bibinfo {author} {\bibfnamefont {G.}~\bibnamefont {Das}}, \bibinfo {author}
  {\bibfnamefont {I.}~\bibnamefont {Maksymov}}, \bibinfo {author}
  {\bibfnamefont {M.}~\bibnamefont {Galli}}, \bibinfo {author} {\bibfnamefont
  {L.}~\bibnamefont {Businaro}}, \bibinfo {author} {\bibfnamefont {L.~C.}\
  \bibnamefont {Andreani}},\ and\ \bibinfo {author} {\bibfnamefont
  {E.}~\bibnamefont {Di~Fabrizio}},\ }\bibfield  {title} {\bibinfo {title} {{A
  {{Hybrid Plasmonic}}-{{Photonic Nanodevice}} for {{Label}}-{{Free Detection}}
  of a {{Few Molecules}}}},\ }\href {https://doi.org/10.1021/nl801112e}
  {\bibfield  {journal} {\bibinfo  {journal} {Nano Lett.}\ }\textbf {\bibinfo
  {volume} {8}},\ \bibinfo {pages} {2321} (\bibinfo {year} {2008})}\BibitemShut
  {NoStop}%
\bibitem [{\citenamefont {Dantham}\ \emph {et~al.}(2013)\citenamefont
  {Dantham}, \citenamefont {Holler}, \citenamefont {Barbre}, \citenamefont
  {Keng}, \citenamefont {Kolchenko},\ and\ \citenamefont
  {Arnold}}]{Dantham2013}%
  \BibitemOpen
  \bibfield  {author} {\bibinfo {author} {\bibfnamefont {V.~R.}\ \bibnamefont
  {Dantham}}, \bibinfo {author} {\bibfnamefont {S.}~\bibnamefont {Holler}},
  \bibinfo {author} {\bibfnamefont {C.}~\bibnamefont {Barbre}}, \bibinfo
  {author} {\bibfnamefont {D.}~\bibnamefont {Keng}}, \bibinfo {author}
  {\bibfnamefont {V.}~\bibnamefont {Kolchenko}},\ and\ \bibinfo {author}
  {\bibfnamefont {S.}~\bibnamefont {Arnold}},\ }\bibfield  {title} {\bibinfo
  {title} {{Label-{{Free Detection}} of {{Single Protein Using}} a
  {{Nanoplasmonic}}-{{Photonic Hybrid Microcavity}}}},\ }\href
  {https://doi.org/10.1021/nl401633y} {\bibfield  {journal} {\bibinfo
  {journal} {Nano Lett.}\ }\textbf {\bibinfo {volume} {13}},\ \bibinfo {pages}
  {3347} (\bibinfo {year} {2013})}\BibitemShut {NoStop}%
\bibitem [{\citenamefont {Conteduca}\ \emph {et~al.}(2016)\citenamefont
  {Conteduca}, \citenamefont {Dell'Olio}, \citenamefont {Innone}, \citenamefont
  {Ciminelli},\ and\ \citenamefont {Armenise}}]{Conteduca2016}%
  \BibitemOpen
  \bibfield  {author} {\bibinfo {author} {\bibfnamefont {D.}~\bibnamefont
  {Conteduca}}, \bibinfo {author} {\bibfnamefont {F.}~\bibnamefont
  {Dell'Olio}}, \bibinfo {author} {\bibfnamefont {F.}~\bibnamefont {Innone}},
  \bibinfo {author} {\bibfnamefont {C.}~\bibnamefont {Ciminelli}},\ and\
  \bibinfo {author} {\bibfnamefont {M.~N.}\ \bibnamefont {Armenise}},\
  }\bibfield  {title} {\bibinfo {title} {{Rigorous Design of an Ultra-High
  {{Q}}/{{V}} Photonic/Plasmonic Cavity to Be Used in Biosensing
  Applications}},\ }\href {https://doi.org/10.1016/j.optlastec.2015.08.016}
  {\bibfield  {journal} {\bibinfo  {journal} {Optics \& Laser Technology}\
  }\textbf {\bibinfo {volume} {77}},\ \bibinfo {pages} {151} (\bibinfo {year}
  {2016})}\BibitemShut {NoStop}%
\bibitem [{\citenamefont {Xavier}\ \emph {et~al.}(2018)\citenamefont {Xavier},
  \citenamefont {Vincent}, \citenamefont {Meder},\ and\ \citenamefont
  {Vollmer}}]{Xavier2018}%
  \BibitemOpen
  \bibfield  {author} {\bibinfo {author} {\bibfnamefont {J.}~\bibnamefont
  {Xavier}}, \bibinfo {author} {\bibfnamefont {S.}~\bibnamefont {Vincent}},
  \bibinfo {author} {\bibfnamefont {F.}~\bibnamefont {Meder}},\ and\ \bibinfo
  {author} {\bibfnamefont {F.}~\bibnamefont {Vollmer}},\ }\bibfield  {title}
  {\bibinfo {title} {{Advances in Optoplasmonic Sensors \textendash{} Combining
  Optical Nano/Microcavities and Photonic Crystals with Plasmonic
  Nanostructures and Nanoparticles}},\ }\href
  {https://doi.org/10.1515/nanoph-2017-0064} {\bibfield  {journal} {\bibinfo
  {journal} {Nanophotonics}\ }\textbf {\bibinfo {volume} {7}},\ \bibinfo
  {pages} {1} (\bibinfo {year} {2018})}\BibitemShut {NoStop}%
\bibitem [{\citenamefont {Hu}\ \emph {et~al.}(2013)\citenamefont {Hu},
  \citenamefont {Li}, \citenamefont {Liu}, \citenamefont {Xiao},\ and\
  \citenamefont {Gong}}]{Hu2013Hybrid}%
  \BibitemOpen
  \bibfield  {author} {\bibinfo {author} {\bibfnamefont {Y.-W.}\ \bibnamefont
  {Hu}}, \bibinfo {author} {\bibfnamefont {B.-B.}\ \bibnamefont {Li}}, \bibinfo
  {author} {\bibfnamefont {Y.-X.}\ \bibnamefont {Liu}}, \bibinfo {author}
  {\bibfnamefont {Y.-F.}\ \bibnamefont {Xiao}},\ and\ \bibinfo {author}
  {\bibfnamefont {Q.}~\bibnamefont {Gong}},\ }\bibfield  {title} {\bibinfo
  {title} {{Hybrid Photonic\textendash Plasmonic Mode for Refractometer and
  Nanoparticle Trapping}},\ }\href
  {https://doi.org/10.1016/j.optcom.2012.11.024} {\bibfield  {journal}
  {\bibinfo  {journal} {Optics Communications}\ }\textbf {\bibinfo {volume}
  {291}},\ \bibinfo {pages} {380} (\bibinfo {year} {2013})}\BibitemShut
  {NoStop}%
\bibitem [{\citenamefont {Hughes}\ \emph {et~al.}(2018)\citenamefont {Hughes},
  \citenamefont {Richter},\ and\ \citenamefont {Knorr}}]{Hughes2018}%
  \BibitemOpen
  \bibfield  {author} {\bibinfo {author} {\bibfnamefont {S.}~\bibnamefont
  {Hughes}}, \bibinfo {author} {\bibfnamefont {M.}~\bibnamefont {Richter}},\
  and\ \bibinfo {author} {\bibfnamefont {A.}~\bibnamefont {Knorr}},\ }\bibfield
   {title} {\bibinfo {title} {{Quantized Pseudomodes for Plasmonic Cavity
  {{QED}}}},\ }\href {https://doi.org/10.1364/OL.43.001834} {\bibfield
  {journal} {\bibinfo  {journal} {Opt. Lett.}\ }\textbf {\bibinfo {volume}
  {43}},\ \bibinfo {pages} {1834} (\bibinfo {year} {2018})}\BibitemShut
  {NoStop}%
\bibitem [{\citenamefont {Franke}\ \emph {et~al.}(2019)\citenamefont {Franke},
  \citenamefont {Hughes}, \citenamefont {Kamandar~Dezfouli}, \citenamefont
  {Kristensen}, \citenamefont {Busch}, \citenamefont {Knorr},\ and\
  \citenamefont {Richter}}]{Franke2019}%
  \BibitemOpen
  \bibfield  {author} {\bibinfo {author} {\bibfnamefont {S.}~\bibnamefont
  {Franke}}, \bibinfo {author} {\bibfnamefont {S.}~\bibnamefont {Hughes}},
  \bibinfo {author} {\bibfnamefont {M.}~\bibnamefont {Kamandar~Dezfouli}},
  \bibinfo {author} {\bibfnamefont {P.~T.}\ \bibnamefont {Kristensen}},
  \bibinfo {author} {\bibfnamefont {K.}~\bibnamefont {Busch}}, \bibinfo
  {author} {\bibfnamefont {A.}~\bibnamefont {Knorr}},\ and\ \bibinfo {author}
  {\bibfnamefont {M.}~\bibnamefont {Richter}},\ }\bibfield  {title} {\bibinfo
  {title} {{Quantization of {{Quasinormal Modes}} for {{Open Cavities}} and
  {{Plasmonic Cavity Quantum Electrodynamics}}}},\ }\href
  {https://doi.org/10.1103/PhysRevLett.122.213901} {\bibfield  {journal}
  {\bibinfo  {journal} {Phys. Rev. Lett.}\ }\textbf {\bibinfo {volume} {122}},\
  \bibinfo {pages} {213901} (\bibinfo {year} {2019})}\BibitemShut {NoStop}%
\bibitem [{\citenamefont {Medina}\ \emph {et~al.}()\citenamefont {Medina},
  \citenamefont {{Garc{\'i}a-Vidal}}, \citenamefont
  {{Fern{\'a}ndez-Dom{\'i}nguez}},\ and\ \citenamefont {Feist}}]{Medina2020}%
  \BibitemOpen
  \bibfield  {author} {\bibinfo {author} {\bibfnamefont {I.}~\bibnamefont
  {Medina}}, \bibinfo {author} {\bibfnamefont {F.~J.}\ \bibnamefont
  {{Garc{\'i}a-Vidal}}}, \bibinfo {author} {\bibfnamefont {A.~I.}\ \bibnamefont
  {{Fern{\'a}ndez-Dom{\'i}nguez}}},\ and\ \bibinfo {author} {\bibfnamefont
  {J.}~\bibnamefont {Feist}},\ }\bibfield  {title} {\bibinfo {title} {{Few-Mode
  {{Field Quantization}} of {{Arbitrary Electromagnetic Spectral
  Densities}}}},\ }\href@noop {} {\ }\Eprint {https://arxiv.org/abs/2008.00349}
  {arXiv:2008.00349} \BibitemShut {NoStop}%
\bibitem [{\citenamefont {Kamandar~Dezfouli}\ and\ \citenamefont
  {Hughes}(2017)}]{KamandarDezfouli2017}%
  \BibitemOpen
  \bibfield  {author} {\bibinfo {author} {\bibfnamefont {M.}~\bibnamefont
  {Kamandar~Dezfouli}}\ and\ \bibinfo {author} {\bibfnamefont {S.}~\bibnamefont
  {Hughes}},\ }\bibfield  {title} {\bibinfo {title} {{Quantum {{Optics Model}}
  of {{Surface}}-{{Enhanced Raman Spectroscopy}} for {{Arbitrarily Shaped
  Plasmonic Resonators}}}},\ }\href
  {https://doi.org/10.1021/acsphotonics.7b00157} {\bibfield  {journal}
  {\bibinfo  {journal} {ACS Photonics}\ }\textbf {\bibinfo {volume} {4}},\
  \bibinfo {pages} {1245} (\bibinfo {year} {2017})}\BibitemShut {NoStop}%
\bibitem [{\citenamefont {Schmidt}\ \emph {et~al.}(2017)\citenamefont
  {Schmidt}, \citenamefont {Esteban}, \citenamefont {Benz}, \citenamefont
  {Baumberg},\ and\ \citenamefont {Aizpurua}}]{Schmidt2017}%
  \BibitemOpen
  \bibfield  {author} {\bibinfo {author} {\bibfnamefont {M.~K.}\ \bibnamefont
  {Schmidt}}, \bibinfo {author} {\bibfnamefont {R.}~\bibnamefont {Esteban}},
  \bibinfo {author} {\bibfnamefont {F.}~\bibnamefont {Benz}}, \bibinfo {author}
  {\bibfnamefont {J.~J.}\ \bibnamefont {Baumberg}},\ and\ \bibinfo {author}
  {\bibfnamefont {J.}~\bibnamefont {Aizpurua}},\ }\bibfield  {title} {\bibinfo
  {title} {{Linking Classical and Molecular Optomechanics Descriptions of
  {{SERS}}}},\ }\href {https://doi.org/10.1039/C7FD00145B} {\bibfield
  {journal} {\bibinfo  {journal} {Faraday Discuss.}\ }\textbf {\bibinfo
  {volume} {205}},\ \bibinfo {pages} {31} (\bibinfo {year} {2017})}\BibitemShut
  {NoStop}%
\bibitem [{\citenamefont {Breuer}\ and\ \citenamefont
  {Petruccione}(2007)}]{Breuer2007}%
  \BibitemOpen
  \bibfield  {author} {\bibinfo {author} {\bibfnamefont {H.-P.}\ \bibnamefont
  {Breuer}}\ and\ \bibinfo {author} {\bibfnamefont {F.}~\bibnamefont
  {Petruccione}},\ }\href
  {https://doi.org/10.1093/acprof:oso/9780199213900.001.0001} {\emph {\bibinfo
  {title} {{The {{Theory}} of {{Open Quantum Systems}}}}}}\ (\bibinfo
  {publisher} {{Oxford University Press}},\ \bibinfo {year} {2007})\BibitemShut
  {NoStop}%
\bibitem [{\citenamefont {Johansson}\ \emph {et~al.}(2012)\citenamefont
  {Johansson}, \citenamefont {Nation},\ and\ \citenamefont
  {Nori}}]{Johansson2012}%
  \BibitemOpen
  \bibfield  {author} {\bibinfo {author} {\bibfnamefont {J.~R.}\ \bibnamefont
  {Johansson}}, \bibinfo {author} {\bibfnamefont {P.~D.}\ \bibnamefont
  {Nation}},\ and\ \bibinfo {author} {\bibfnamefont {F.}~\bibnamefont {Nori}},\
  }\bibfield  {title} {\bibinfo {title} {{{{QuTiP}}: {{An}} Open-Source
  {{Python}} Framework for the Dynamics of Open Quantum Systems}},\ }\href
  {https://doi.org/10.1016/j.cpc.2012.02.021} {\bibfield  {journal} {\bibinfo
  {journal} {Comput. Phys. Commun.}\ }\textbf {\bibinfo {volume} {183}},\
  \bibinfo {pages} {1760} (\bibinfo {year} {2012})}\BibitemShut {NoStop}%
\bibitem [{\citenamefont {Johansson}\ \emph {et~al.}(2013)\citenamefont
  {Johansson}, \citenamefont {Nation},\ and\ \citenamefont
  {Nori}}]{Johansson2013}%
  \BibitemOpen
  \bibfield  {author} {\bibinfo {author} {\bibfnamefont {J.~R.}\ \bibnamefont
  {Johansson}}, \bibinfo {author} {\bibfnamefont {P.~D.}\ \bibnamefont
  {Nation}},\ and\ \bibinfo {author} {\bibfnamefont {F.}~\bibnamefont {Nori}},\
  }\bibfield  {title} {\bibinfo {title} {{{{QuTiP}} 2: {{A Python}} Framework
  for the Dynamics of Open Quantum Systems}},\ }\href
  {https://doi.org/10.1016/j.cpc.2012.11.019} {\bibfield  {journal} {\bibinfo
  {journal} {Comput. Phys. Commun.}\ }\textbf {\bibinfo {volume} {184}},\
  \bibinfo {pages} {1234} (\bibinfo {year} {2013})}\BibitemShut {NoStop}%
\bibitem [{\citenamefont {Hunter}(2007)}]{Hunter2007}%
  \BibitemOpen
  \bibfield  {author} {\bibinfo {author} {\bibfnamefont {J.~D.}\ \bibnamefont
  {Hunter}},\ }\bibfield  {title} {\bibinfo {title} {{Matplotlib: {{A 2D
  Graphics Environment}}}},\ }\href {https://doi.org/10.1109/MCSE.2007.55}
  {\bibfield  {journal} {\bibinfo  {journal} {Comput. Sci. Eng.}\ }\textbf
  {\bibinfo {volume} {9}},\ \bibinfo {pages} {90} (\bibinfo {year}
  {2007})}\BibitemShut {NoStop}%
\bibitem [{\citenamefont {Caswell}\ \emph {et~al.}(2020)\citenamefont
  {Caswell}, \citenamefont {Droettboom}, \citenamefont {Lee}, \citenamefont
  {Hunter}, \citenamefont {Firing}, \citenamefont {Stansby}, \citenamefont
  {Klymak}, \citenamefont {Hoffmann}, \citenamefont {{Sales de Andrade}},
  \citenamefont {Varoquaux}, \citenamefont {Hedegaard~Nielsen}, \citenamefont
  {Root}, \citenamefont {Elson}, \citenamefont {May}, \citenamefont {Dale},
  \citenamefont {Lee}, \citenamefont {Sepp{\"a}nen}, \citenamefont {McDougall},
  \citenamefont {Straw}, \citenamefont {Hobson}, \citenamefont {Gohlke},
  \citenamefont {Yu}, \citenamefont {Ma}, \citenamefont {Vincent},
  \citenamefont {Silvester}, \citenamefont {Moad}, \citenamefont {Kniazev},
  \citenamefont {Ivanov}, \citenamefont {Ernest},\ and\ \citenamefont
  {Katins}}]{Caswell2020}%
  \BibitemOpen
  \bibfield  {author} {\bibinfo {author} {\bibfnamefont {T.~A.}\ \bibnamefont
  {Caswell}}, \bibinfo {author} {\bibfnamefont {M.}~\bibnamefont {Droettboom}},
  \bibinfo {author} {\bibfnamefont {A.}~\bibnamefont {Lee}}, \bibinfo {author}
  {\bibfnamefont {J.}~\bibnamefont {Hunter}}, \bibinfo {author} {\bibfnamefont
  {E.}~\bibnamefont {Firing}}, \bibinfo {author} {\bibfnamefont
  {D.}~\bibnamefont {Stansby}}, \bibinfo {author} {\bibfnamefont
  {J.}~\bibnamefont {Klymak}}, \bibinfo {author} {\bibfnamefont
  {T.}~\bibnamefont {Hoffmann}}, \bibinfo {author} {\bibfnamefont
  {E.}~\bibnamefont {{Sales de Andrade}}}, \bibinfo {author} {\bibfnamefont
  {N.}~\bibnamefont {Varoquaux}}, \bibinfo {author} {\bibfnamefont
  {J.}~\bibnamefont {Hedegaard~Nielsen}}, \bibinfo {author} {\bibfnamefont
  {B.}~\bibnamefont {Root}}, \bibinfo {author} {\bibfnamefont {P.}~\bibnamefont
  {Elson}}, \bibinfo {author} {\bibfnamefont {R.}~\bibnamefont {May}}, \bibinfo
  {author} {\bibfnamefont {D.}~\bibnamefont {Dale}}, \bibinfo {author}
  {\bibfnamefont {J.-J.}\ \bibnamefont {Lee}}, \bibinfo {author} {\bibfnamefont
  {J.~K.}\ \bibnamefont {Sepp{\"a}nen}}, \bibinfo {author} {\bibfnamefont
  {D.}~\bibnamefont {McDougall}}, \bibinfo {author} {\bibfnamefont
  {A.}~\bibnamefont {Straw}}, \bibinfo {author} {\bibfnamefont
  {P.}~\bibnamefont {Hobson}}, \bibinfo {author} {\bibfnamefont
  {C.}~\bibnamefont {Gohlke}}, \bibinfo {author} {\bibfnamefont {T.~S.}\
  \bibnamefont {Yu}}, \bibinfo {author} {\bibfnamefont {E.}~\bibnamefont {Ma}},
  \bibinfo {author} {\bibfnamefont {A.~F.}\ \bibnamefont {Vincent}}, \bibinfo
  {author} {\bibfnamefont {S.}~\bibnamefont {Silvester}}, \bibinfo {author}
  {\bibfnamefont {C.}~\bibnamefont {Moad}}, \bibinfo {author} {\bibfnamefont
  {N.}~\bibnamefont {Kniazev}}, \bibinfo {author} {\bibfnamefont
  {P.}~\bibnamefont {Ivanov}}, \bibinfo {author} {\bibfnamefont
  {E.}~\bibnamefont {Ernest}},\ and\ \bibinfo {author} {\bibfnamefont
  {J.}~\bibnamefont {Katins}},\ }\href {https://doi.org/10.5281/zenodo.3714460}
  {\bibinfo {title} {{Matplotlib v3.2.1}}},\ \bibinfo {howpublished} {Zenodo}
  (\bibinfo {year} {2020})\BibitemShut {NoStop}%
\bibitem [{\citenamefont {Gardiner}\ and\ \citenamefont
  {Zoller}(2004)}]{Gardiner2004}%
  \BibitemOpen
  \bibfield  {author} {\bibinfo {author} {\bibfnamefont {C.~W.}\ \bibnamefont
  {Gardiner}}\ and\ \bibinfo {author} {\bibfnamefont {P.}~\bibnamefont
  {Zoller}},\ }\href@noop {} {\emph {\bibinfo {title} {{Quantum {{Noise}}: {{A
  Handbook}} of {{Markovian}} and {{Non}}-{{Markovian Quantum Stochastic
  Methods}} with {{Applications}} to {{Quantum Optics}}}}}}\ (\bibinfo
  {publisher} {{Springer Berlin Heidelberg}},\ \bibinfo {year}
  {2004})\BibitemShut {NoStop}%
\end{thebibliography}%

\end{document}